\documentclass[journal]{IEEEtran}
\usepackage{amsmath,amssymb,amsfonts}
\usepackage[ruled,linesnumbered]{algorithm2e}
\usepackage{array}
\usepackage[caption=false,font=footnotesize,labelfont=rm,textfont=rm]{subfig}
\usepackage{textcomp}
\usepackage{stfloats}
\usepackage{multirow}
\usepackage{makecell}
\usepackage{url}
\usepackage{verbatim}
\usepackage{graphicx}
\usepackage{cite}
\usepackage{hyperref}
\usepackage{booktabs}
\usepackage{wrapfig}
\usepackage{threeparttable}
\usepackage{bbding}
\usepackage[commandnameprefix=always]{changes}

\hypersetup{hypertex=true,
            colorlinks=true,
            linkcolor=blue,
            anchorcolor=blue,
            citecolor=blue}
\begin{document}

\title{TRIS-HAR: Transmissive Reconfigurable Intelligent Surfaces-assisted Cognitive Wireless Human Activity Recognition Using State Space Models}
\author{Junshuo Liu,~\IEEEmembership{Student Member,~IEEE},
        Yunlong Huang,
        Xin Shi,~\IEEEmembership{Member,~IEEE},
        Rujing Xiong,~\IEEEmembership{Student Member,~IEEE},
        Jianan Zhang,
        Tiebin Mi,~\IEEEmembership{Member,~IEEE},
        Robert C. Qiu,~\IEEEmembership{Fellow,~IEEE}

\thanks{Manuscript received April xx, 202x; revised August xx, 202x. This work was supported in part by the Nation Natural Science Foundation of China under Grant No.12141107, in part by the Guangxi Science and Technology Project (AB21196034), and in part by the Interdisciplinary Research Program of HUST, 2023JCYJ012.  \textit{(Corresponding author: Tiebin Mi.)}}

\thanks{J. Liu, Y. Huang, R. Xiong, J. Zhang, T. Mi, and R. Qiu are with the School of Electronic Information and Communications, Huazhong University of Science and Technology, Wuhan 430074, China (e-mail: junshuo\_liu@hust.edu.cn; huangyunlong@hust.edu.cn; rujing@hust.edu.cn; zhangjn@hust.edu.cn; mitiebin@hust.edu.cn; caiming@hust.edu.cn).}

\thanks{X. Shi is with the Energy Internet Research Institute, Tsinghua University, Beijing 100085, China (e-mail: xinshi\_bjcy@163.com).}
}

\markboth{Journal of \LaTeX\ Class Files,~Vol.~14, No.~8, August~2021}%
{Shell \MakeLowercase{\textit{et al.}}: A Sample Article Using IEEEtran.cls for IEEE Journals}


\maketitle

\begin{abstract}
Human activity recognition (HAR) using radio frequency (RF) signals has garnered considerable attention for its applications in smart environments. However, traditional systems often struggle with limited independent channels between transmitters and receivers, multipath fading, and environmental noise, which particularly degrades performance in through-the-wall scenarios. In this paper, we present a transmissive reconfigurable intelligent surface (TRIS)-assisted through-the-wall human activity recognition (TRIS-HAR) system. The system employs TRIS technology to actively reshape wireless signal propagation, creating multiple independent paths to enhance signal clarity and improve recognition accuracy in complex indoor settings. Additionally, we propose the Human intelligence Mamba (HiMamba), an advanced state space model that captures temporal and frequency-based information for precise activity recognition. HiMamba achieves state-of-the-art performance on two public datasets, demonstrating superior accuracy. Extensive experiments indicate that the TRIS-HAR system improves recognition performance from 85.00\% to 98.06\% in laboratory conditions and maintains high performance across various environments. This approach offers a robust solution for enhancing RF-based HAR, with promising applications in smart home and elderly care systems.
\end{abstract}


\section{Introduction}
\IEEEPARstart{H}{uman} activity recognition (HAR) using radio frequency (RF) signals has emerged as a pivotal technology in smart home automation, elderly care, and emergency response systems \cite{wang2018deep,liu2019wireless,zhang2020device,shi2022fedrfid}. Traditional RF-based HAR systems employ channel state information (CSI) to detect human movements by analyzing signal propagation paths and capturing changes in amplitude and phase due to environmental interactions. Systems like CARM \cite{wang2017device} and TSHNN \cite{huang2024tshnn} demonstrate the effectiveness of CSI in recognizing human activities by analyzing signal distortions caused by these movements. However, the complexity and unpredictability of wireless environments significantly impact the accuracy and flexibility of activity recognition. This is primarily due to undesired multipath fading, environmental noise, and the limited number of independent channels available between transmitters and receivers in conventional HAR systems \cite{qian2018widar2}.

The accuracy of activity recognition further deteriorates in through-the-wall scenarios, where signals must penetrate obstacles such as walls. Walls attenuate and distort signals, exacerbating multipath fading and environmental noise, which substantially impedes accurate activity recognition in adjacent rooms or behind closed doors \cite{adib20143d,wang2019survey}. Through-the-wall recognition can enhance security and surveillance in critical areas where a direct line of sight is not possible. Consequently, there is an urgent need to develop more robust technologies that enhance signal clarity and improve recognition performance in these challenging environments.

Recently, reconfigurable intelligent surface (RIS) technology has significantly advanced the capability to actively customize propagation channels, thereby creating a favorable electromagnetic environment to address challenges in complex indoor settings, such as multipath fading and environmental noise \cite{huang2019reconfigurable,elmossallamy2020reconfigurable,tang2020wireless}. Reflective RIS (RRIS) primarily redirects incoming signals to optimize their paths, enhancing signal strength where line-of-sight communication is obstructed. However, this technology faces limitations in scenarios requiring signal penetration through obstacles like walls, as both the transmitter and receiver must be situated on the same side of the RRIS, imposing geographical constraints on the physical topology \cite{wang2023doppler}.

To address this issue, the concept of transmissive RIS (TRIS) is gaining increasing attention \cite{tang2023transmissive,li2023towards,li2024transmissive}. Unlike RRIS, TRIS enables signals to pass through, forming directional beams that manipulate the phase and amplitude of the signal. This capability is crucial for enhancing the accuracy of HAR in through-the-wall scenarios by allowing RF signals to penetrate obstacles like walls and reshape as they traverse complex environments \cite{zhou2017short,hu2020reconfigurable}. Consequently, TRIS technology improves the clarity and reliability of signals through walls, enhancing the detectability of subtle human movements across barriers. Each independent path created by the TRIS carries distinct and valuable information about human activities, thereby significantly boosting HAR accuracy, especially in scenarios where RRIS schemes may fail to provide sufficient spatial information \cite{basar2019wireless,zhang2022toward}.

In this study, we develop a transmissive reconfigurable intelligent surface-assisted through-the-wall human activity recognition (TRIS-HAR) system. This system leverages TRIS technology to enhance the performance of RF-based HAR in complex environments. By precisely manipulating the phase and amplitude of RF signals, the TRIS-HAR system mitigates prevalent issues such as multipath fading and improves signal clarity in multi-room or wall-dense settings. Additionally, the capability of TRIS to generate multiple independent and reconfigurable paths enriches the spatial information, which is crucial for detecting and recognizing human activities. This innovative integration not only preserves high signal integrity across barriers but also improves the overall accuracy and reliability of activity recognition in challenging indoor scenarios.

Furthermore, we propose Human intelligence Mamba (HiMamba), a novel model designed to process and analyze the unique characteristics of sensing data. HiMamba extends traditional state space models by incorporating advanced techniques to effectively capture the temporal and spatial dynamics of human activities. It employs a dual-stream framework to process both temporal and frequency information, allowing for a comprehensive understanding of complex signal variations. This architecture surpasses existing models on two publicly available datasets, demonstrating superior ability in modeling and recognizing human activities from CSI data.

Finally, we validate the effectiveness of the TRIS-HAR system through extensive experiments on our datasets. The integration of TRIS significantly improves activity recognition accuracy, especially in through-the-wall scenarios, with an increase in accuracy from 85.00\% to 98.06\%. The system also demonstrates robustness across various environments, maintaining high performance in both controlled and real-world settings. These results confirm the TRIS-HAR system's adaptability and effectiveness in enhancing RF-based HAR under challenging conditions.

The key contributions of our research are delineated as follows:
\begin{itemize}
    \item We develop a novel through-the-wall human activity recognition system utilizing the TRIS, which markedly enhances RF-based HAR performance by improving signal propagation through walls and mitigating environmental noise effects. To the best of our knowledge, this is the first attempt to leverage the TRIS for through-the-wall human activity recognition.
    
    \item We propose the HiMamba model, which extends traditional state space models to effectively process CSI data for human activity recognition. HiMamba employs a dual-stream framework to capture both temporal and frequency information. Due to its parallelization, HiMamba exhibits better computational efficiency compared with sequential models.
    
    \item The proposed model achieves state-of-the-art performance while maintaining a lightweight structure. Compared with the best-performing baselines, our model achieves superior performance with only 3.14\% of the parameter size, significantly reducing training and inference time in practice.

    \item We conduct extensive experiments on two public datasets and three real-world datasets to validate the advantage of our proposed method over state-of-the-art baselines for RF-based HAR tasks. On the real-world datasets, the proposed model maintains high accuracy and demonstrates robustness across various environmental settings.
\end{itemize}

The rest of this paper is as follows: Section II reviews related work on RF-based HAR and RIS technologies. Section III discusses preliminary concepts, including the TRIS and its configuration for enhancing signals. Section IV describes the TRIS-HAR system, detailing its architecture, hardware setup, data preprocessing methods, and the proposed state space model. Section V presents the experiments and results, demonstrating the TRIS-HAR system's effectiveness and robustness. Section VI concludes the paper and outlines future research directions.

\section{Related Work}
\subsection{Human Activity Recognition Using RF signals}
Recent advancements in leveraging channel state information for human activity recognition have demonstrated its efficacy in capturing and analyzing fine-grained signal variations. CSI-based HAR utilizes alterations in RF signals induced by human movements to detect and classify various activities. Wang \textit{et al.} \cite{wang2017device} demonstrated that CSI could distinguish activities such as walking, running, and boxing by analyzing signal changes caused by human movement. This seminal work established the foundation for using CSI in activity differentiation. Building upon this, Li \textit{et al.} \cite{li2019wi} introduced Wi-Motion, which enhanced this capability by integrating both amplitude and phase information from CSI, achieving precise activity classification even in complex environments, thereby broadening CSI's applicability in diverse settings. Lu \textit{et al.} \cite{lu2022cehar} further advanced the field with CeHAR, which employed Convolutional Neural Networks (CNNs) to dynamically fuse CSI features, resulting in improved recognition accuracy and efficiency. These developments underscore CSI's potential as a robust tool for non-intrusive HAR, effectively capturing a range of human activities through detailed signal analysis.

However, these works primarily focus on scenarios without significant obstacles such as walls. Our TRIS-HAR system addresses this gap by integrating the TRIS to enhance RF-based HAR, significantly improving recognition accuracy in through-the-wall scenarios. This advancement demonstrates the potential of TRIS in overcoming the limitations of traditional RF-based systems in complex environments.

\subsection{Reconfigurable Intelligent Surfaces in Wireless Sensing}
Reconfigurable intelligent surfaces have emerged as a transformative technology in wireless sensing, dynamically manipulating electromagnetic waves to shape and control the propagation environment, thereby overcoming traditional limitations in wireless communication and sensing.

Wu and Zhang \cite{wu2019towards} provided foundational insights into how RIS can create adaptable wireless environments by actively directing signals, enhancing signal quality and coverage. Their work illustrated the fundamental principles of RIS technology and its potential applications. Building on this framework, Hu \textit{et al.} \cite{hu2020reconfigurable} developed a system for RIS-based RF sensing, optimizing configurations to improve accuracy in posture recognition. This study highlighted the effectiveness of RIS in fine-tuning electromagnetic environments for enhanced sensing capabilities. Wang \textit{et al.} \cite{wang2023integrated} explored RIS-assisted backscatter systems, proposing algorithms for the joint optimization of sensing and communication. Their work demonstrated significant improvements in overall system performance, illustrating RIS's potential to mitigate challenges such as multipath interference and poor signal environments.

Recent advancements in stacked intelligent metasurfaces (SIMs), a next-generation evolution of RIS, have further extended the potential of metasurface technology. SIMs enable advanced wave-domain signal processing, facilitating functionalities like multi-input multi-output (MIMO) precoding, multi-user interference cancellation, and direction-of-arrival (DOA) estimation directly in the electromagnetic domain. Liu \textit{et al.} \cite{liu2024stacked} discussed the applications and challenges of SIM technology for wireless sensing and communication, while An \textit{et al.} \cite{an2024two} demonstrated its efficacy in two-dimensional DOA estimation. Moreover, SIM-based MIMO transceiver designs offer substantial improvements in computational efficiency and sensing accuracy by leveraging wave-based processing capabilities \cite{an2024stacked}. These works highlight the growing importance of SIMs in enabling low-latency, power-efficient sensing and communication systems.

While these studies demonstrate the potential of RIS and SIM in various applications, our work uniquely applies the TRIS to human activity recognition. By employing a TRIS to create multiple independent signal paths, the TRIS-HAR system enhances signal clarity and improves recognition accuracy in through-the-wall scenarios, serving as a practical use of RIS in HAR.

\subsection{Advances in Deep Learning for Wireless Sensing}
Deep learning has advanced human activity recognition using CSI by effectively modeling spatial, temporal, and contextual patterns. Several architectures, such as CNNs, Long Short-Term Memory (LSTM) networks, and Transformers, have been utilized to capture the complex dynamics within CSI data. Lin \textit{et al.} \cite{lin2022human} employed CNNs to extract spatial features from CSI, achieving high-accuracy activity recognition through local pattern extraction. This approach demonstrated the capability of CNNs to leverage spatial information from wireless signals. Addressing temporal dependencies, Zhang \textit{et al.} \cite{zhang2020data} implemented Dense-LSTM networks, which enhanced the recognition of sequential activities in CSI data by effectively capturing the temporal dynamics of human activities. Furthermore, Xu \textit{et al.} \cite{xu2022time} developed a sparse-connected LSTM-based deep learning framework for accurate CSI prediction in RIS-assisted multi-user multiple-input single-output networks, thereby improving the efficiency of CSI acquisition in time-varying channels. More recently, Transformer models have emerged as a powerful tool in HAR. Luo \textit{et al.} \cite{luo2024vision} applied Vision Transformers to model long-range dependencies and complex interactions in CSI data, demonstrating their superior performance in recognizing intricate activity patterns.

In recent years, a notable trend is the emergence of artificial intelligence-generated content (AIGC). Wang \textit{et al.} \cite{wang2024generative} investigated the application of generative AI models, including generative adversarial networks (GANs) and diffusion models, to enhance RF sensing in Internet of Things (IoT) systems. Their focus was on overcoming challenges such as data augmentation, missing modalities, and multi-modal fusion, aimed at improving the scalability and robustness of wireless sensing applications. In another study, Wang \textit{et al.} \cite{wang2024aigc} introduced RFID-ACCDM, a conditional diffusion model that synthesizes high-fidelity RFID data for HAR. They demonstrated that the generated data could outperform models trained with real data. Furthermore, Wang \textit{et al.} \cite{wang2024generativeAI} proposed G-HFD, a generative AI-assisted human flow detection system that utilizes CSI for velocity and acceleration estimation. This system employs a unified weighted conditional diffusion model for denoising and fine-grained flow detection, demonstrating high accuracy in detecting subflows and their sizes in practical communication environments.

These deep learning models have shown remarkable capabilities in HAR by effectively analyzing CSI data. Our TRIS-HAR system further advances this field by introducing the HiMamba model, an advanced state space model that processes enhanced CSI data from the TRIS. By capturing both temporal and frequency features, HiMamba achieves state-of-the-art performance on public datasets, showcasing the synergy between deep learning techniques and RIS technology for robust HAR.

In summary, while significant progress has been made in HAR using CSI, RIS technology, and deep learning models, our TRIS-HAR system uniquely combines these advancements to address the challenges of through-the-wall scenarios, demonstrating superior accuracy and robustness in complex indoor environments.

\section{Preliminary}
\subsection{Transmissive Reconfigurable Intelligent Surfaces}
\begin{figure}[tbp]
\centerline{\includegraphics[width=0.8\columnwidth]{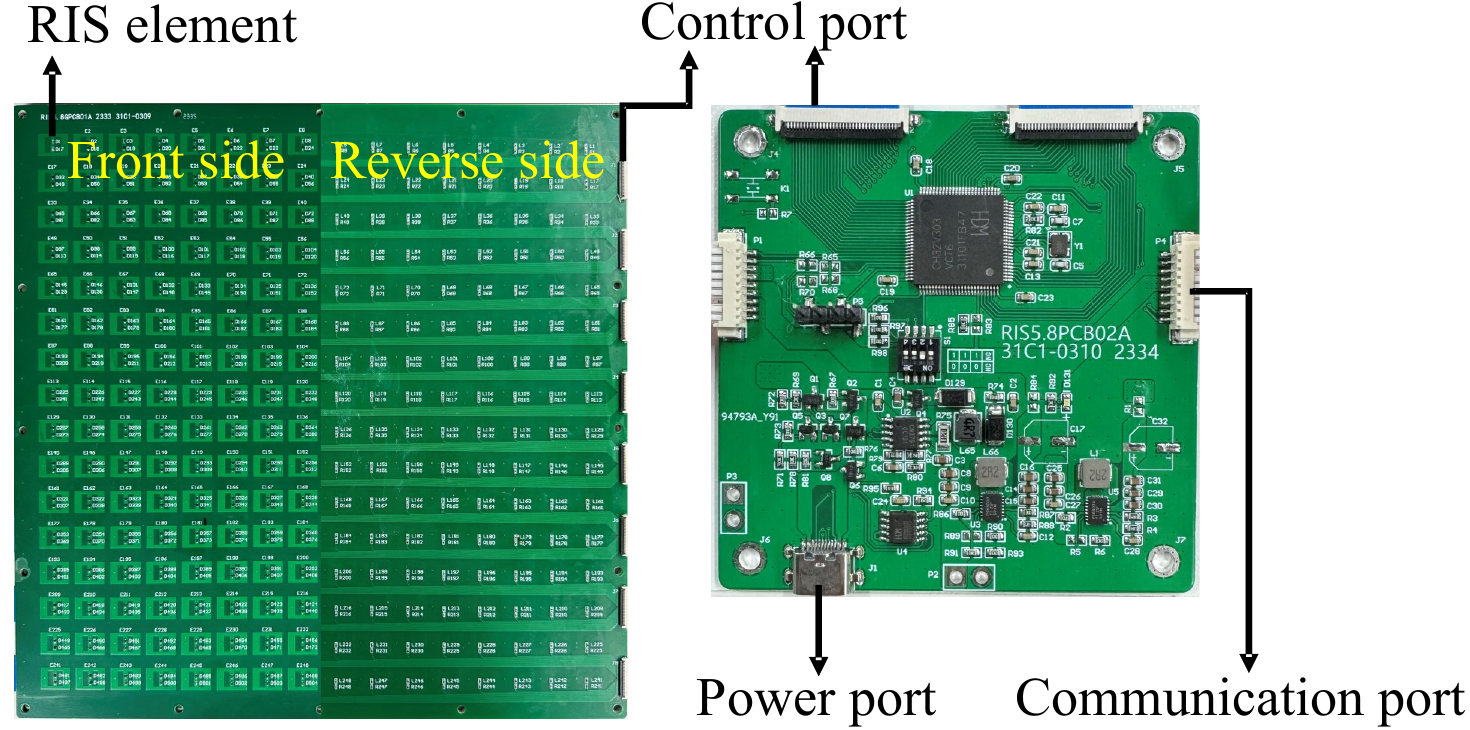}}
\caption{Photography of the 1-bit TRIS prototype (left side), and the logic circuit control board (right side).}
\label{F1}
\end{figure}

We depict a prototype of a 1-bit transmissive reconfigurable intelligent surface, operating at a center frequency of 5.8 GHz, as shown in Fig.~\ref{F1}. It features a uniform planar array (UPA) consisting of a $16 \times 16$ elements, with overall dimensions of $31 \times 31 \, \text{cm}^2$. The design's primary aim is to enable phase reconfigurability by allowing individual control over each element.

To facilitate this, the prototype incorporates a network of 196 biasing lines, each connected to a single element to control it via DC voltages applied across two PIN diodes. The operation of these lines is managed by four logic circuit control boards, which are linked through various connectors. 
Each element includes two PIN diodes in an anti-symmetric configuration, allowing for a switch between two bias states that correspond to different operational modes. Specifically, a '0' state occurs when the first PIN diode is off and the second is on, and a '1' state occurs with the reverse configuration. This setup enables a phase shift of approximately 180 degrees at 5.8 GHz, which is crucial for the RIS's phase reconfigurability.

\subsection{Optimization of TRIS Configurations}
A single-antenna transmitter (Tx) communicates with a single-antenna receiver (Rx), facilitated by a TRIS consisting of $L = M \times N$ elements. The received signal $y \in \mathbb{C}$ is influenced by the TRIS's phase shifts, as expressed by:
\begin{equation}
y = \boldsymbol{h}_1^{H} \boldsymbol{W} \boldsymbol{h}_2 x + n,
\end{equation}
where $\boldsymbol{h}_1^{H} \in \mathbb{C}^{L}$ and $\boldsymbol{h}_2 \in \mathbb{C}^{L}$ represent the channel vectors from the TRIS to the Rx and from the Tx to the TRIS, respectively. The transmitted signal is denoted as $x \in \mathbb{C}$, and $n \sim \mathcal{CN}(0,\sigma^2_n)$ signifies additive white Gaussian noise with variance $\sigma^2_n$. The configuration matrix $\boldsymbol{W} = \mathrm{diag}(e^{j\beta_1}, e^{j\beta_2},...,e^{j\beta_L}) \in \mathbb{C}^{L \times L}$ encapsulates the TRIS's phase shifts, with each transmission coefficient $e^{j\beta_n}$ reflecting a phase shift $\beta_n \in [-\pi, \pi]$ at the respective TRIS element.

The signal power at the receiver is proportional to the squared magnitude of the combined channel gain, denoted as $\lvert \boldsymbol{h}_1^{H} \boldsymbol{W} \boldsymbol{h}_2 \rvert ^2$, which can be manipulated through strategic configurations of $\boldsymbol{W}$. {{Our prototype employs quantized phase shifts of either $-\pi / 2$ or $+\pi / 2$, optimizing signal power incrementally through a configuration optimization algorithm. This algorithm adjusts phase shifts either row-by-row or column-by-column, as described in Algorithm~\ref{algorithm1}.}}

\begin{algorithm}[tbp]
    \caption{{Configuration optimization algorithm.}}
    \DontPrintSemicolon
    \label{algorithm1}
    \SetKwInOut{Input}{Input}\SetKwInOut{Output}{Output}
    \Input{Initial received signal power, $p_0$.}
    \Output{Optimized transmission coefficients matrix, $\boldsymbol{\Phi}_{M+N}$.}
    \BlankLine
    Initialize $\boldsymbol{\Phi}_0 \in \mathbb{C}^{M \times N}$ with phase shifts of $-\pi / 2$;\\
    \For{$m \leftarrow 1$ \KwTo $M$}{
    Modify $\boldsymbol{\Phi}_m$ by inverting the phase of the $m$-th row of $\boldsymbol{\Phi}_{m-1}$;\\
    Measure power $p_m$;\\
    \If{$p_m < p_{m-1}$}{
    $\boldsymbol{\Phi}_m \leftarrow \boldsymbol{\Phi}_{m-1}$;
    }
    }
    \For{$n \leftarrow 1$ \KwTo $N$}{
    Modify $\boldsymbol{\Phi}_{M+n}$ by inverting the phase of the $n$-th column of $\boldsymbol{\Phi}_{M+n-1}$;\\
    Measure power $p_{M+n}$;\\
    \If{$p_{M+n} < p_{M+n-1}$}{
    $\boldsymbol{\Phi}_{M+n} \leftarrow \boldsymbol{\Phi}_{M+n-1}$;
    }
    }
\end{algorithm}

This algorithm iteratively inverts the phase of specific rows or columns within the TRIS. If a phase modification improves the power of the received signal relative to the previous configuration, the adjustment is retained. The optimization process is managed by a logic circuit control board. Transmission coefficients are represented by an $M \times N$ complex matrix $\boldsymbol{\Phi}_t$, where $\boldsymbol{W}$ is realized through the diagonalization of $\boldsymbol{\Phi}_t$. Each element $(m, n)$ in $\boldsymbol{\Phi}_t$ corresponds to the transmission coefficient of respective TRIS element located at the $m$-th row and $n$-th column. The matrix is structured as follows:
\begin{equation}
\boldsymbol{\Phi}_t = [\boldsymbol{c}_{t,1}, \boldsymbol{c}_{t,2},\dots,\boldsymbol{c}_{t,N}] = [\boldsymbol{r}_{t,1},\boldsymbol{r}_{t,2},\dots,\boldsymbol{r}_{t,M}]^T,
\end{equation}
where $\boldsymbol{c}_{t,n} \in \mathbb{C}^{M}$ and $\boldsymbol{r}_{t,m} \in \mathbb{C}^{N}$ are the transmission coefficients of the $n$-th column and $m$-th row, respectively. The average power of the received signal at time $t$ is denoted as $p_t$.

\subsection{Wall Attenuation and Signal Enhancement}
Path loss (PL) is defined as the ratio of the effective transmitted power to the received power, accounting for system losses and gains from amplifiers and antennas. For link budget analysis, the path loss relative to free space (FS) at one meter is used as a reference \cite{rappaport2010wireless}:
\begin{equation}
    \begin{aligned}
        P_R & =  P_T + G_T + G_R - [\text{Path Loss w.r.t. 1 m FS}] \nonumber \\
            & + 20 \log_{10} \left ( \frac{\lambda}{4 \pi d}\right ),
    \end{aligned}
\end{equation}
where $\lambda$ represents the wavelength, and $G_T$ and $G_R$ are the antenna gains of the transmitter and receiver, respectively, expressed in decibels (dB). The powers of the transmitter ($P_T$) and receiver ($P_R$) are measured in dBm.

The model for distance-dependent path loss is given by:
\begin{equation}
\overline{PL}(d) [\text{dB}] = PL(d_0) [\text{dB}] + 10 n \log_{10} \left ( \frac{d}{d_0} \right ),
\end{equation}
where $\overline{PL}(d)$ denotes the average path loss in dB at a distance $d$, $PL(d_0)$ is the path loss at a reference distance of 1 m, and $n$ is the path loss exponent that indicates the increase in path loss with the distance between the transmitter and receiver \cite{rappaport2010wireless}.

Advanced propagation models include partition-dependent attenuation, starting with a base free space path loss where $n = 2$ and adding attenuation from obstructions:
\begin{equation}
P_R = P_T + G_T + G_R + 20\log_{10} \left ( \frac{\lambda}{4 \pi d} \right ) - \sum_{i=1}^{N} \alpha_i,
\label{PLM}
\end{equation}
where $\alpha_i$ measures the attenuation from the $i$-th obstruction \cite{durgin1998measurements}.

\begin{figure}[tbp]
\centerline{\includegraphics[width=1.0\columnwidth]{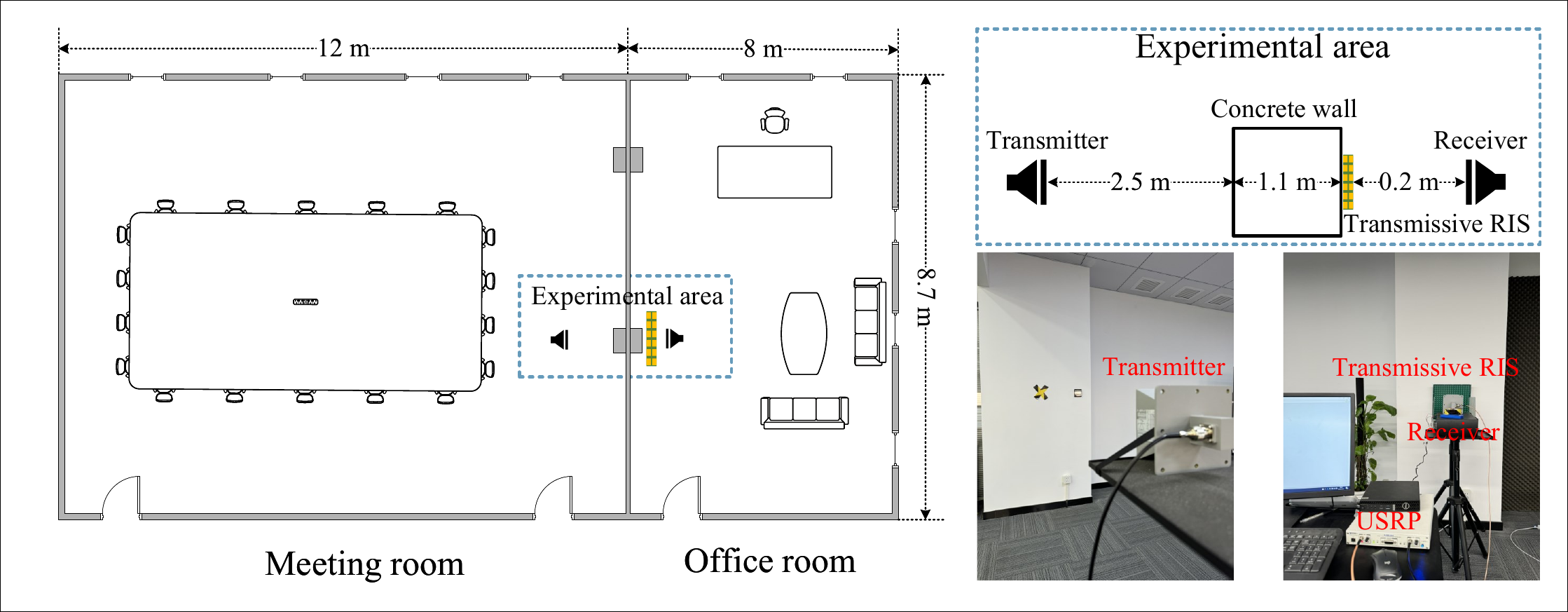}}
\caption{Floor plans of the experimental environment with concrete wall.}
\label{F2}
\end{figure}

In our laboratory experiment, detailed in Fig.~\ref{F2}, we use the National Instruments (NI) Universal Software Radio Peripheral (USRP)-2954R and Laboratory Virtual Instrument Engineering Workbench (LabVIEW) software. The setup includes a 3.8-meter propagation distance, with Cable 1 and Cable 2 measuring 3 meters and 10 meters, respectively. Horn antennas (Model: HD-58SGH15N) for both transmission and reception are connected via Cable 2 and Cable 1, respectively, resulting in a total cable length of 13 meters and an average signal attenuation of 1.27 dB/m at 5.8 GHz. The parameters for path loss measurements are detailed in Table~\ref{table1}.

\begin{table}[htbp]
    \centering
    \caption{Configurations in channel measurements.}
    \label{table1}
    \setlength{\tabcolsep}{4mm}
    \begin{tabular}{ll}
    \toprule
    Parameters                 & Narrowband measurement
    \\ 
    \midrule
    Frequency                  & 5.8 GHz
    \\
    Probing signal             & Continuous wave (CW) 
    \\
    Type of Tx/Rx antenna      & Directional horn antenna 
    \\
    Transmitted power          & 17 dBm          
    \\
    Amplifier power            & 14 dB @ 5.8 GHz     
    \\
    Antenna gain               & 15.8 dBi @ 5.8 GHz    
    \\
    Height of Tx/Rx antenna    & 1 m             
    \\
    Dimension of TRIS          & $16 \times 16$       
    \\
    \bottomrule
    \end{tabular}
\end{table}

To assess the impact of concrete wall obstructions on signal attenuation, it is essential to precisely determine the wall's electromagnetic properties within the experimental framework. Relative permittivity ($\varepsilon_r$) and conductivity ($\sigma$) significantly influence signal attenuation \cite{pozar2011microwave}. Relative permittivity, represented as $\varepsilon_r(f)$, is the ratio of the material's permittivity to the electric permittivity of a vacuum, defined as:
\begin{equation}
\varepsilon_r(f) = \frac{\varepsilon(f)}{\varepsilon_0},
\end{equation}
where $\varepsilon(f)$ is the material's complex frequency-dependent permittivity, and $\varepsilon_0$ is the permittivity of vacuum.

Radio wave propagation experiences attenuation when passing through materials. While the general formula for the dielectric constant is intricate, it can be simplified at two extremes: the dielectric limit ($\sigma \rightarrow 0$) and the good conductor limit ($\sigma \rightarrow \infty$) \cite{rudd2014building,series2015effects}:
\begin{equation}
\alpha = 1636 \frac{\sigma}{\sqrt{\varepsilon_r^{'}}},
\label{Attenuation}
\end{equation}
where $\varepsilon_r^{'}$ is the real part of $\varepsilon_r$, and $\alpha$ denotes the attenuation in dB/m for a specific material. For concrete at 5.8 GHz, the parameters range from $\varepsilon_r^{'}=3.58$ to $5.50$ and $\sigma=0.11 \, \text{S/m}$ \cite{cuinas2000comparison,cuinas2001measuring,cuinas2002permittivity,cuinas2007modelling,ferreira2014review}.

\begin{figure}[tbp]
    \centering
    \subfloat[]{
    \label{F3-a}
    \includegraphics[width=0.47\columnwidth]{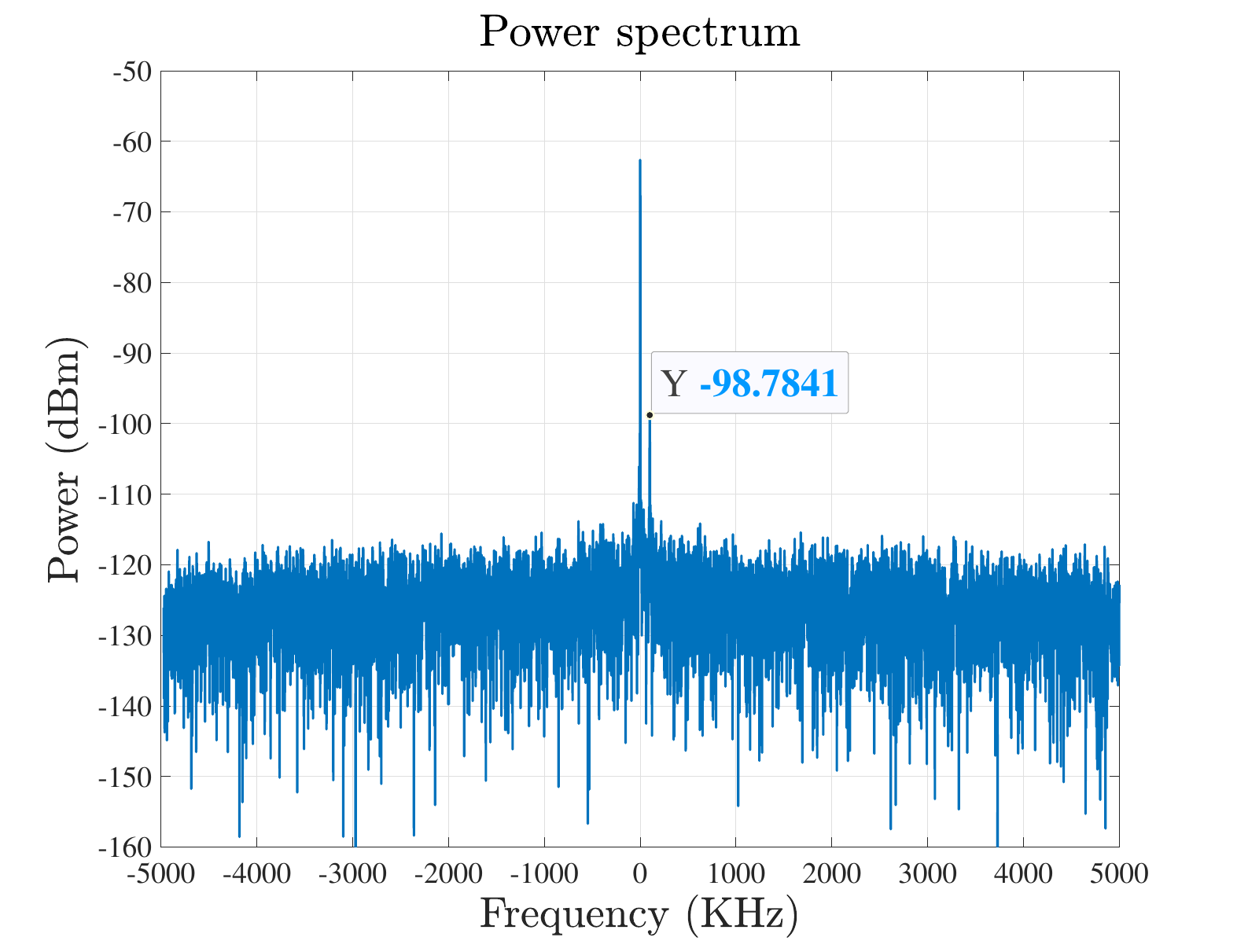}
    }
    \subfloat[]{
    \label{F3-b}
    \includegraphics[width=0.47\columnwidth]{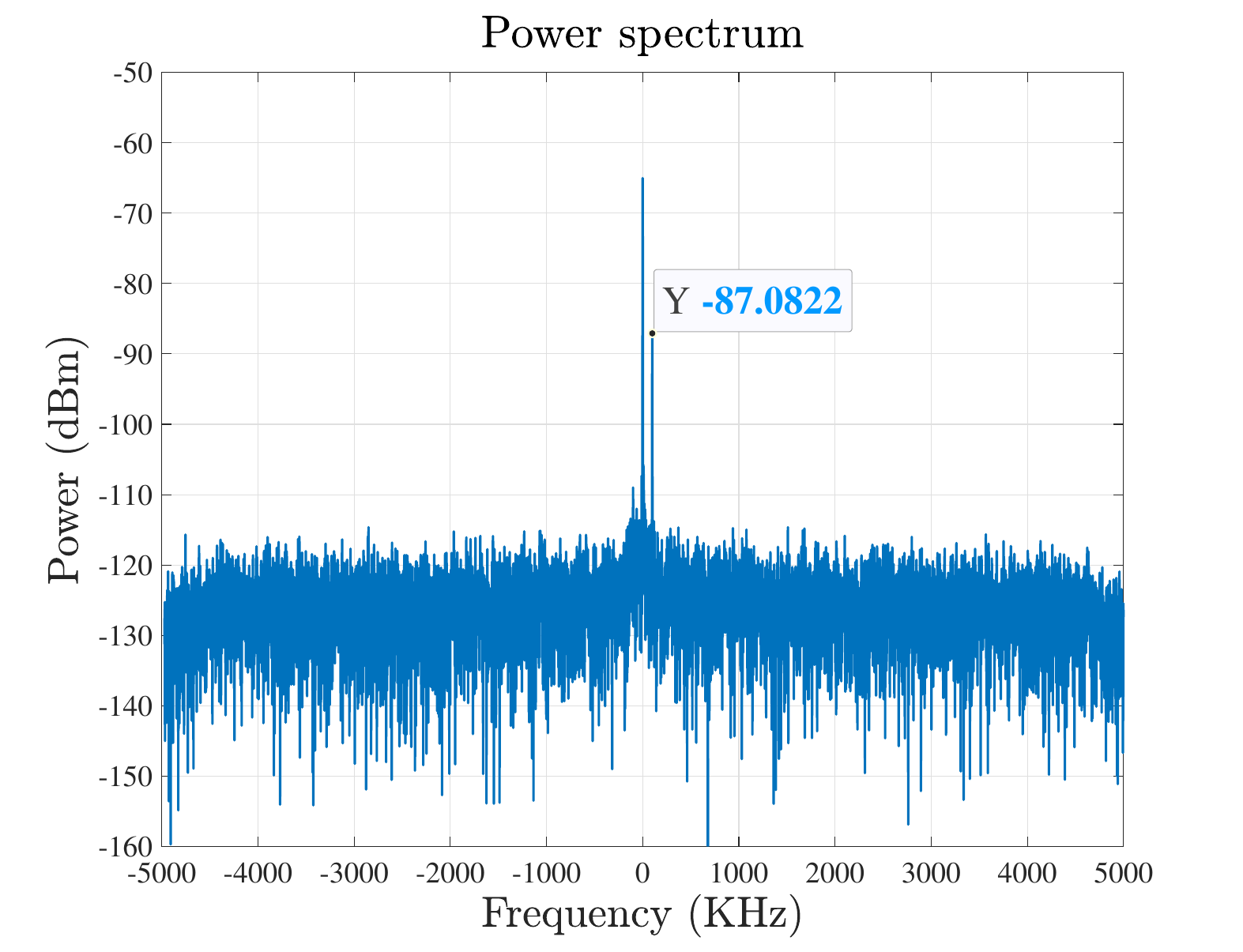}
    }
    \caption{The spectrum of the TRIS test: (a) without TRIS; (b) with TRIS.}
    \label{F3}
\end{figure}

Leveraging these values and the formulas outlined, we calculate the receiver power to be -98.52 dBm. During our validation experiments, with the carrier frequency set as 5.8 GHz and a bandwidth of 100 kHz, the received signal power is quantified by averaging over 8192 data points. Without the TRIS, the received signal power is -98.78 dBm, as shown in Fig.~\ref{F3}~\subref{F3-a}. However, after integrating the TRIS and optimizing the transmission coefficients matrix $\boldsymbol{\Phi}$ via the configuration optimization algorithm, we achieve a peak signal strength of -87.08 dBm, as shown in Fig.~\ref{F3}~\subref{F3-b}. This corresponds to an 11.7 dBi array gain at the operational frequency, demonstrating the TRIS's effectiveness in enhancing signal strength and overcoming transmission obstructions.

\subsection{Channel State Information}
\begin{figure}[tbp]
    \centering
    \subfloat[Without RIS.]{
    \label{CFRNoRIS}
    \includegraphics[width=0.9\columnwidth]{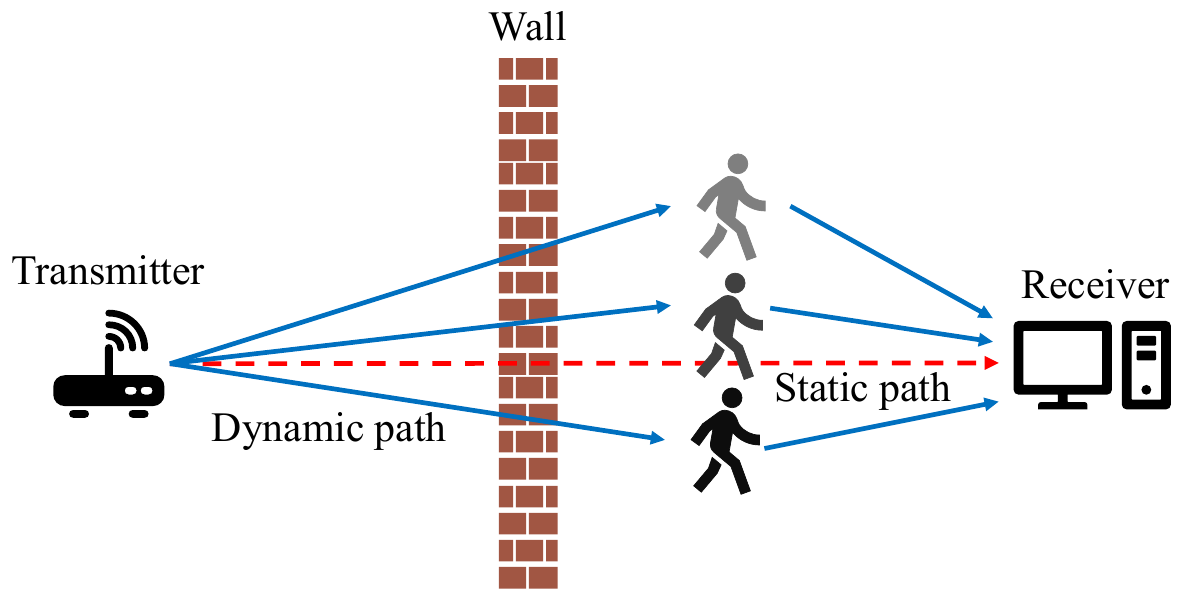}}
    \\
    \subfloat[With RIS.]{
    \label{CFRRIS}
    \includegraphics[width=0.9\columnwidth]{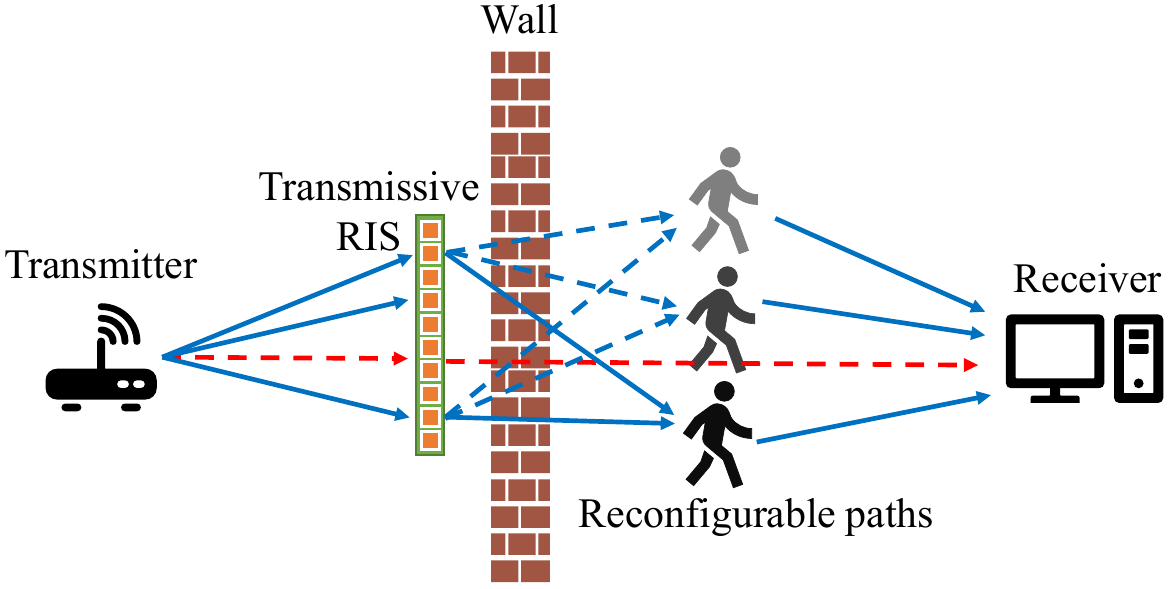}}
    \caption{Signal variation diagram while an object moves in through-the-wall scenarios.}
    \label{CFR}
\end{figure}

In wireless communications, CSI describes the signal propagation from the transmitter to the receiver, capturing the effects of reflection, diffraction, and scattering in the physical environment. In real-world through-the-wall scenarios, CSI measures the combined effects of multipath channels, encompassing both static and dynamic components, as depicted in Fig.~\ref{CFR}~\subref{CFRNoRIS}. The static components of CSI represent fixed paths where the signal penetrates through the wall and interacts with the environment's stable features, such as walls, floors, and furniture. The dynamic components capture variations in signal paths due to moving objects, such as humans, introducing time-varying delays and phase shifts. These variations are crucial for sensing human activity in the environment.

Mathematically, the overall CSI can be represented as:
\begin{equation*}
    \begin{aligned}
        H(f, t) & = \sum_{m=1}^{M} \beta_{wall} \alpha_m(f) e^{j (\phi_m + \phi_{wall} - 2 \pi f (\tau_m + \tau_{wall}) )} \\
                & + \sum_{n=1}^{N} \beta_{wall} \alpha_n(f, t) e^{j (\phi_n + \phi_{wall} - 2\pi f (\tau_n(t) + \tau_{wall}))},
    \end{aligned}
\end{equation*}
where $\beta_{wall}$ is the attenuation factor due to the wall. $\alpha_m(f)$ and $\alpha_n(f, t)$ are the attenuation coefficients for the $m$-th static path and $n$-th dynamic path, respectively. $\phi_{wall}$ represents the phase shift introduced by the wall, while $\phi_m$ and $\phi_n$ are the phase shifts for the static and dynamic paths. $\tau_{wall}$ is the additional delay caused by the wall, and $\tau_m$ and $\tau_n(t)$ denote the delays for the static and dynamic paths.

With the introduction of the TRIS, the CSI can be significantly enhanced by reconfiguring the static and dynamic paths, as shown in Fig.~\ref{CFR}~\subref{CFRRIS}, thereby compensating for the wall's effects and optimizing signal propagation \cite{hu2020programmable}. The enhanced CSI can be expressed as:
\begin{equation}
    \begin{aligned}
        H_{RIS}(f, t) & = \sum_{m=1}^{M} \beta_{wall} \alpha_m(f) \beta_{RIS, m} (t) e^{j\omega_m} \\
        & + \sum_{n=1}^{N} \beta_{wall} \alpha_n(f, t) \beta_{RIS, n} (t) e^{j\omega_n}, \\
    \end{aligned}
\end{equation}
where
\begin{small}
\begin{equation*}
    \begin{aligned}
        \omega_m & = \phi_m + \theta_{RIS, m}(t) + \phi_{wall} - 2 \pi f (\tau_m + \tau_{wall} + \tau_{RIS,m}(t)), \\
        \omega_n & = \phi_n + \theta_{RIS, n}(t) + \phi_{wall} - 2\pi f (\tau_n(t) + \tau_{wall} + \tau_{RIS,n}(t)).
    \end{aligned}
\end{equation*}
\end{small}
Here, $\beta_{RIS, m}(t)$ and $\beta_{RIS, n}(t)$ represent the TRIS-induced gain for static and dynamic paths, respectively. $\theta_{RIS, m}(t)$ and $\theta_{RIS, n}(t)$ indicate the phase adjustments introduced by the TRIS. $\tau_{RIS, m}(t)$ and $\tau_{RIS, n}(t)$ are the additional delays introduced by the TRIS. By adjusting the gains $\beta_{RIS}$ and phase shifts $\theta_{RIS}$ for both static and dynamic paths, TRIS helps to counteract the adverse effects of walls and other obstacles, creating a more favorable propagation environment. This enhances the overall CSI, thereby improving the accuracy of through-the-wall sensing and communication.

In conclusion, integrating TRIS into wireless systems transforms the channel environment by reconfiguring the static and dynamic components of the CSI. This capability allows for improved control over signal propagation, mitigating the challenges posed by through-the-wall scenarios, and achieving more reliable and precise wireless communication and sensing.

\section{TRIS-HAR System}

\begin{figure*}[tbp]
    \centerline{\includegraphics[width=1.9\columnwidth]{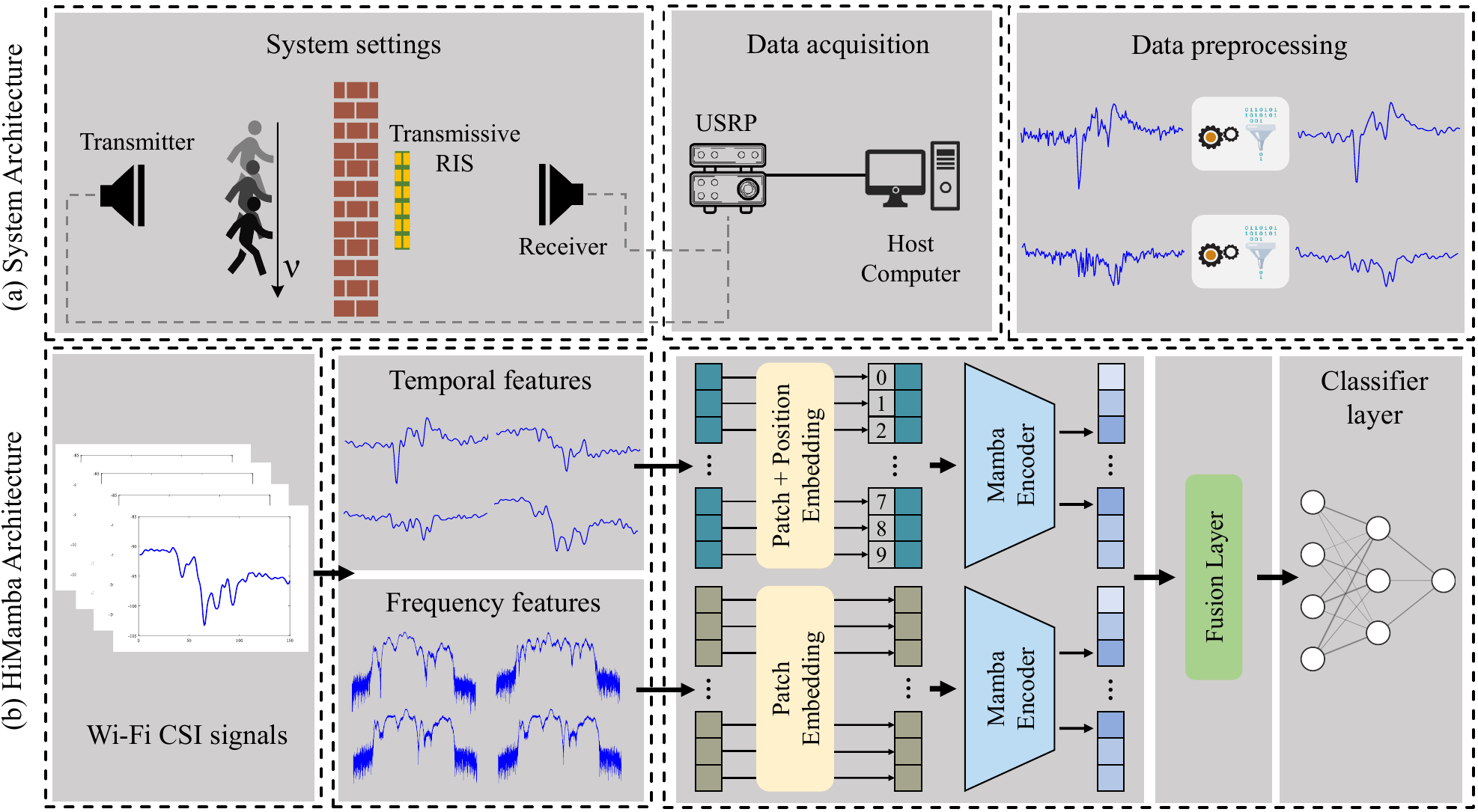}}
    \caption{The overall architecture of the TRIS-HAR system.} 
    \label{TRIS-HAR-Arch}
\end{figure*}

\subsection{System Architecture and Test Settings}
The architecture of the proposed TRIS-HAR system is illustrated in Fig.~\ref{TRIS-HAR-Arch}. It includes three main components: the data acquisition module, the feature extraction module, and the activity recognition module. To demonstrate its benefits in practical applications, we implement our system using a software-defined radio (SDR) platform, specifically the NI USRP-2954R device. Practical experiments are conducted to verify the effectiveness of our system design and proposed algorithms. 


The system operates at a frequency of 5.8 GHz with a bandwidth of 160 MHz, employing linear frequency modulation (LFM) chirp signals. It consists of a USRP for signal transmission and reception, a TRIS, and both transmitter and receiver horn antennas. The USRP, which includes a transmitter and a receiver, manages the signal transmission and reception. The TRIS, composed of $16 \times 16$ passive antenna units, enhances signal strength and mitigates transmission obstructions. Each unit's state is controlled by diodes and managed by a host computer.

\begin{figure}[tbp]
    \centering
    \subfloat[Laboratory.]{
    \label{F4-a}
    \includegraphics[width=0.375\columnwidth]{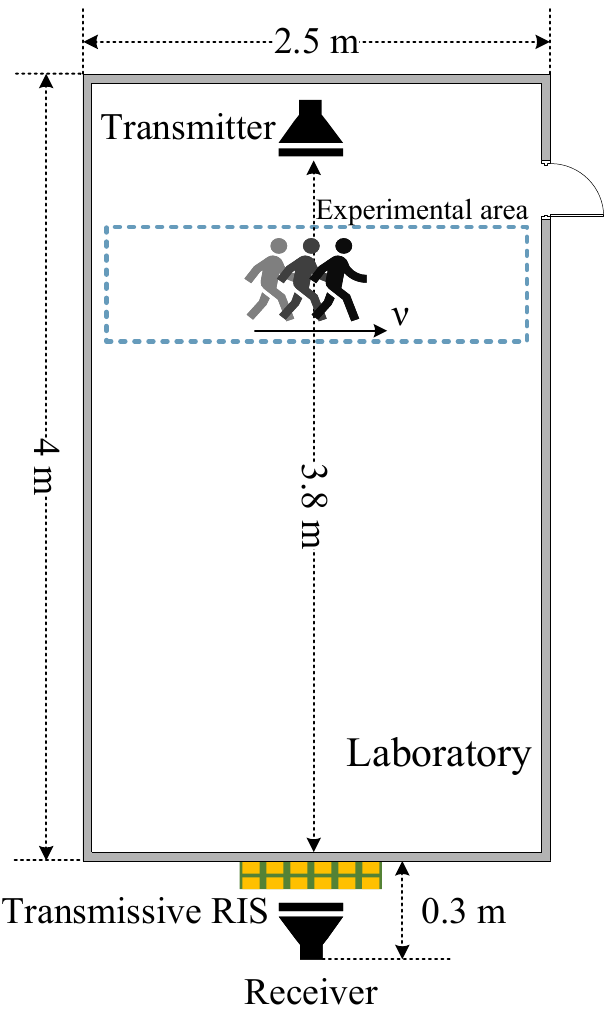}}
    \subfloat[Office.]{
    \label{F4-b}
    \includegraphics[width=0.59\columnwidth]{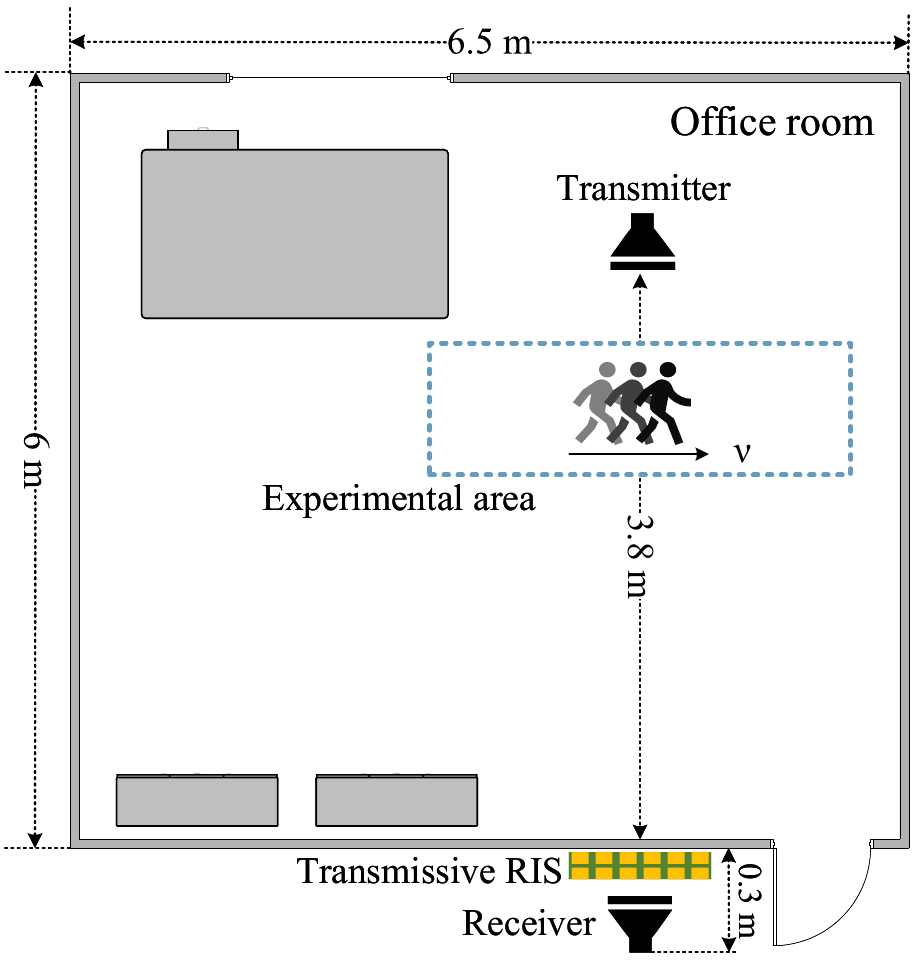}}
    \\
    \subfloat[Meeting room.]{
    \label{F4-c}
    \includegraphics[width=0.98\columnwidth]{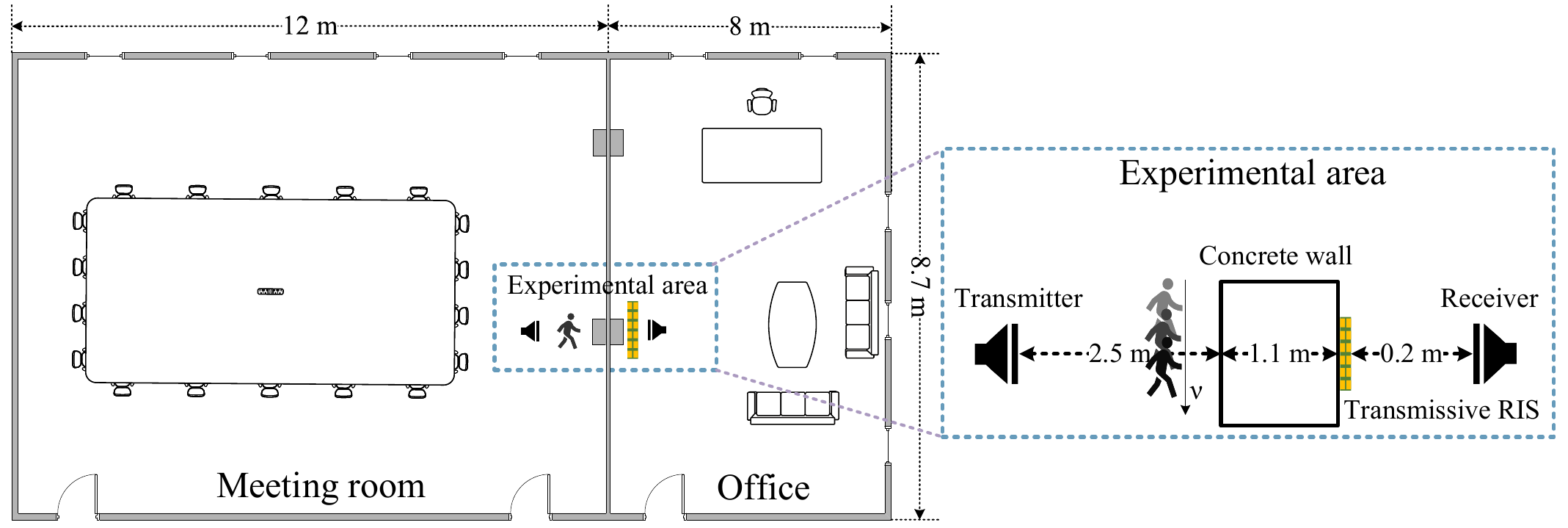}}
    \caption{Floor plans of three experimental environments: (a) Environment 1: Laboratory; (b) Environment 2: Office; (c) Environment 3: Meeting room.}
    \label{F4}
\end{figure}

Performance evaluation of the TRIS-HAR system is conducted in three different environments, as shown in Fig.~\ref{F4}. Each environment features concrete walls and varies in layout:
\begin{itemize}
    \item Environment 1: A laboratory and adjacent lobby, measuring 4 m by 2.5 m.
    
    \item Environment 2: An office room and adjoining corridor, with the office measuring 6.5 m by 6 m.
    
    \item Environment 3: A meeting room and neighboring office room, where the meeting room measures 12 m by 8.7 m, and the office room 8.7 m by 8 m.
\end{itemize}

\subsection{Data Acquisition}
Data is collected at a sampling rate of 50 Hz over a 3-second interval. Each LFM frame comprises 8192 samples with a sampling rate of 160 MHz, resulting in CSI data frames with dimensions of $8192 \times 150$. We record six human activities, which include kicking, picking up, sitting down, standing, standing up, and walking, performed by six subjects in a laboratory setting. Each subject repeats each activity 50 times, yielding 300 samples per activity. To assess the efficacy of TRIS-enabled human activity recognition, CSI data are collected in the laboratory both with and without the aid of the TRIS. Details of these scenarios are provided in Table~\ref{selfdataset}.

\begin{table*}[]
    \centering
    \caption{Details and statistics of three scenarios.}
    \label{selfdataset}
    \setlength{\tabcolsep}{3.5mm}
    \begin{tabular}{ccccc}
    \toprule
    Scenarios                           & Laboratory & Laboratory & Office & Meeting room \\
    \midrule
    TRIS-assisted           & \Checkmark & \XSolidBrush & \Checkmark & \Checkmark            \\
    Number of activity categories       &  6     &   6   &     6    &    6        \\
    Number of subjects (males, females) &  6 (5, 1)    &   6 (5, 1)   &    10 (7, 3)   &   5 (5, 0)           \\
    Number of samples per category      &  50    &   50  &    50    & 100         \\
    \bottomrule   
    \end{tabular}
\end{table*}

\subsection{Data Preprocessing}
\begin{figure}[tbp]
    \centering
    \subfloat[Original CSI stream.]{
    \label{OriginalCSI}
    \includegraphics[width=0.47\columnwidth]{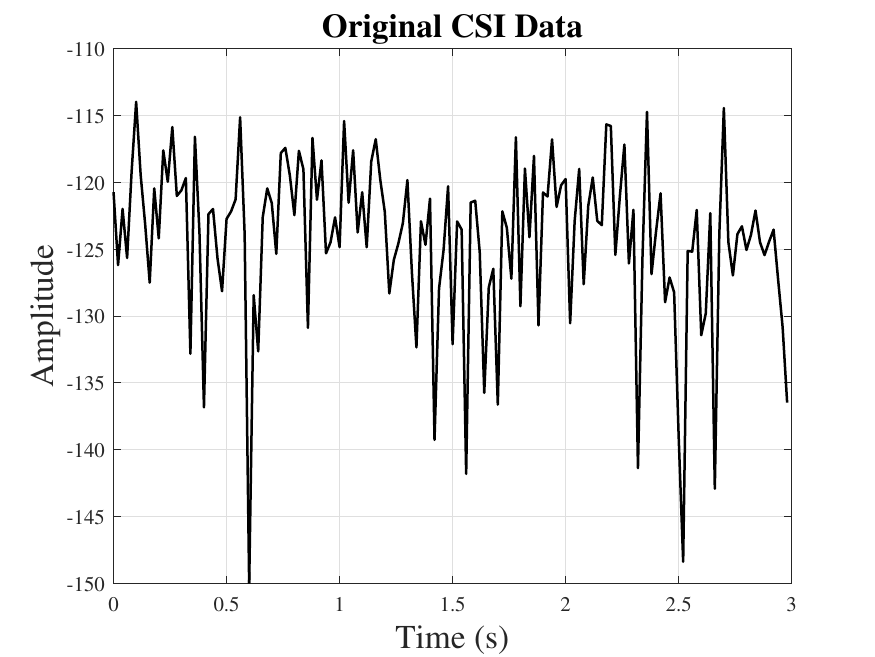}}
    \subfloat[Butterworth low-pass filter.]{
    \label{FilteredCSI}
    \includegraphics[width=0.47\columnwidth]{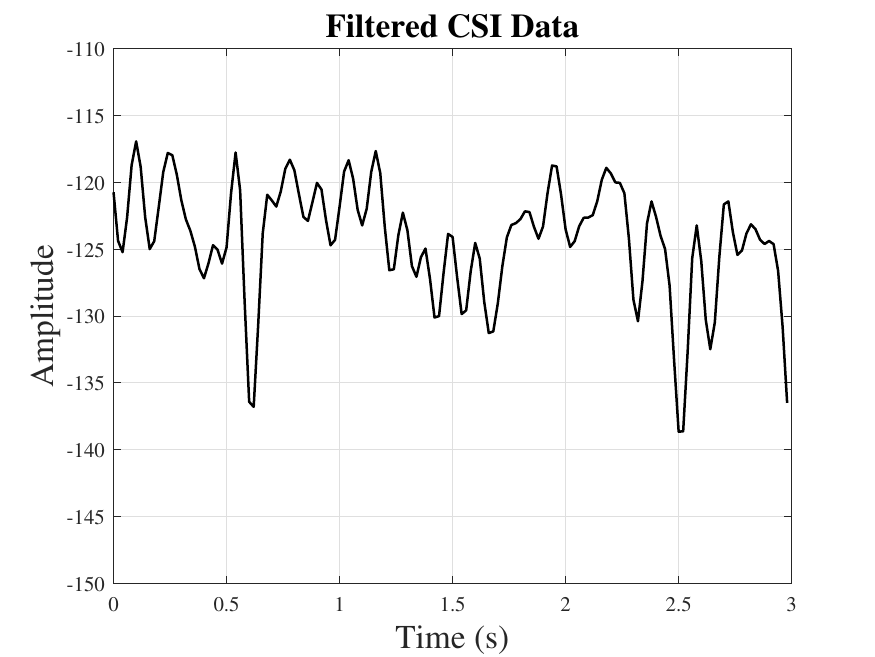}}
    \caption{Denoising the time-series of CSI values.}
    \label{CSIPreprocessing}
\end{figure}

The collection of CSI is significantly impacted by complex wireless propagation and environmental factors, introducing substantial noise into the CSI amplitude values associated with various human activities. This noise is particularly evident as high-frequency noise within each CSI stream, which is unrelated to the intended human activities \cite{chen2023cross}. As depicted in Fig.~\ref{CSIPreprocessing}~\subref{OriginalCSI}, the raw CSI stream segment shows a high level of noise.

To address this issue, we refer to prior works \cite{wang2016rt} and \cite{fridolfsson2019effects} and employ a low-pass filtering method that effectively removes high-frequency noise while preserving relevant low-frequency activity-related signals. Specifically, we select the Butterworth filter for its well-documented ability to provide a maximally flat frequency response in the passband. Empirical evidence from previous studies, such as \cite{wang2016rt}, suggests that human motion-induced CSI variations predominantly occur at lower frequencies, typically below 10 Hz. Accordingly, we set the cutoff frequency of the Butterworth filter at 10 Hz. This setting achieves a balance between retaining meaningful low-frequency components of the CSI data and filtering out irrelevant high-frequency noise. Fig.~\ref{CSIPreprocessing} illustrates the raw and filtered CSI streams, demonstrating the filter's effectiveness in reducing noise.

The application of the Butterworth filter significantly reduces high-frequency noise, resulting in a cleaner and more interpretable signal. This filtered CSI data forms a more robust foundation for subsequent feature extraction and classification in human activity recognition.

\subsection{Data Analysis: HiMamba}
In this subsection, we propose the Human intelligence Mamba model. The goal of HiMamba is to apply state space models (SSMs) for human activity recognition.

\subsubsection{S4 Model and Mamba}
The structured state space sequence (S4) model \cite{gu2021efficiently}, inspired by continuous systems, maps a one-dimensional function or sequence $x(t) \in \mathbb{R} \rightarrow y(t) \in \mathbb{R}$ through an implicit latent state $\boldsymbol{h}(t) \in \mathbb{R}^N$. It provides a robust framework for modeling physical systems, particularly linear time-invariant (LTI) systems, utilizing $\boldsymbol{A} \in \mathbb{R}^{N \times N}$ as the transition matrix, with $\boldsymbol{B} \in \mathbb{R}^{N \times 1}$ and $\boldsymbol{C} \in \mathbb{R}^{1 \times N}$ serving as projection vectors. The model is described by the following state-space representation:
\begin{equation}
    \begin{aligned}\label{SSM}
        \boldsymbol{h}^{'}(t) & = \boldsymbol{A}\boldsymbol{h}(t) + \boldsymbol{B} x(t), \\
        y(t) & = \boldsymbol{C}\boldsymbol{h}(t).
    \end{aligned}
\end{equation}

To apply this model to a discrete input sequence $\boldsymbol{x} = \{x_0, x_1, \cdots\}$ rather than a continuous function $x(t)$, Equation~\eqref{SSM} must be discretized using a step size $\Delta$, which defines the input's resolution. Inputs $x_k$ are considered samples of an underlying continuous signal $x(t)$, where $x_k = x(k\Delta)$. The transformation commonly employs zero-order hold (ZOH), defined as:
\begin{equation}
    \begin{aligned}\label{ZOH}
        \boldsymbol{\bar{A}} & = \exp(\Delta \boldsymbol{A}), \\
        \boldsymbol{\bar{B}} & = (\Delta \boldsymbol{A})^{-1} (\exp(\Delta \boldsymbol{A} - \boldsymbol{I})) \cdot \Delta \boldsymbol{B}.
    \end{aligned}
\end{equation}

After converting the continuous parameters $(\Delta, \boldsymbol{A}, \boldsymbol{B}, \boldsymbol{C})$ to discrete equivalents $(\boldsymbol{\bar{A}}, \boldsymbol{\bar{B}}, \boldsymbol{C})$ via~\eqref{ZOH}, the discrete SSM is represented as:
\begin{equation}\label{SSMref}
    \begin{aligned}
        \boldsymbol{h}_k & = \boldsymbol{\bar{A}} \boldsymbol{h}_{k-1} + \boldsymbol{\bar{B}} x_k, \\
        y_k & = \boldsymbol{C} \boldsymbol{h}_k.
    \end{aligned}
\end{equation}

This can be vectorized into a convolution with an explict formula for the SSM convolution kernel $\boldsymbol{\bar{K}} \in \mathbb{R}^L$.
\begin{equation}
    \begin{aligned}
        \boldsymbol{\bar{K}} & = (\boldsymbol{\bar{C}} \boldsymbol{\bar{B}}, \boldsymbol{\bar{C}} \boldsymbol{\bar{A}} \boldsymbol{\bar{B}}, \dots, \boldsymbol{\bar{C}} \boldsymbol{\bar{A}}^{L-1} \boldsymbol{\bar{B}}), \\
        y & = \boldsymbol{\bar{K}} \times \boldsymbol{x} .
    \end{aligned}
\end{equation}

However, the S4 model's performance is limited on certain tasks due to its linear time-invariance. Unlike S4, the Mamba model adapts parameters $(\boldsymbol{B}, \boldsymbol{C})$ and the step size $\Delta$, based on the input sequence length and batch size, allowing for dynamic and selective state updates. This adaptation makes Mamba a time-variant model, enhancing its ability to capture and represent temporal features effectively. Additionally, Mamba enables parallelization using the parallel scan algorithm \cite{martin2017parallelizing}, forming the selective scan structured state space sequence (S6) model \cite{gu2023mamba}.

\begin{figure}[tbp]
    \centerline{\includegraphics[width=0.95\columnwidth]{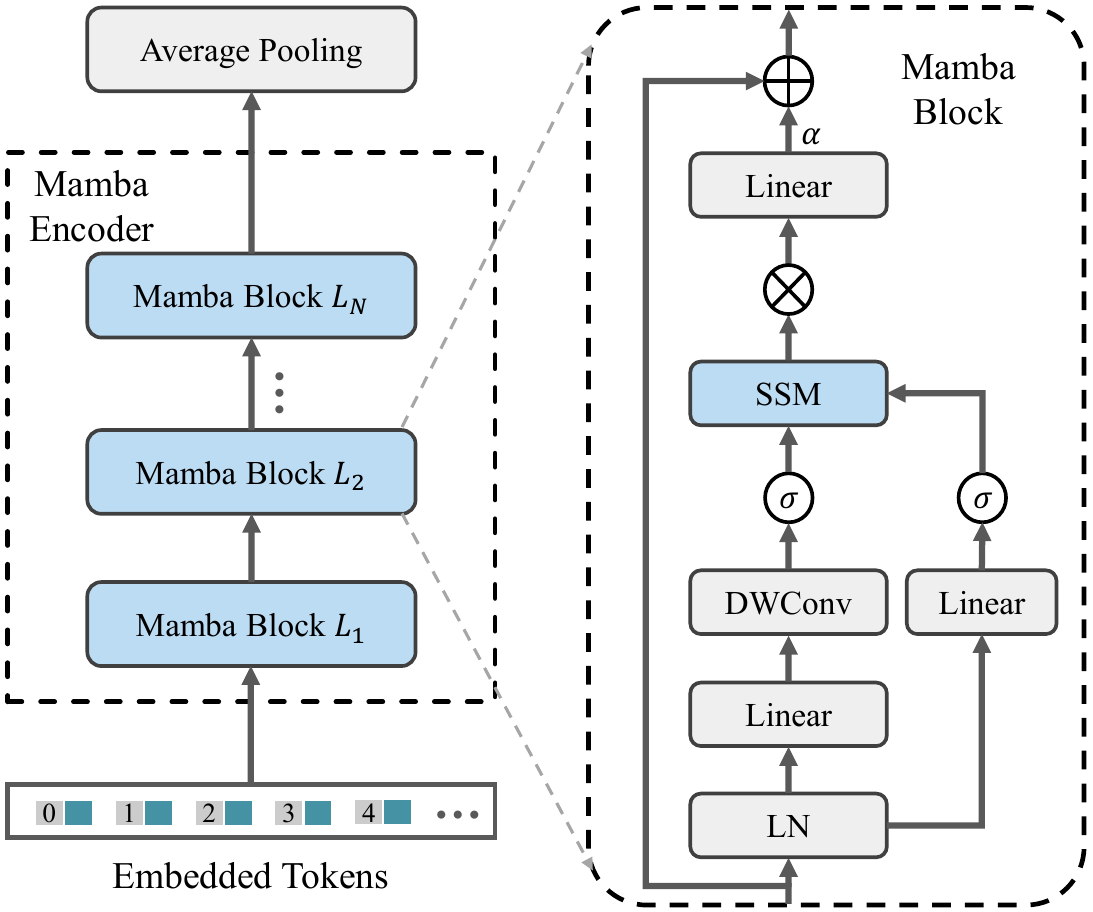}}
    \caption{The structure of the Mamba encoder.}
    \label{MambaEncoder}
\end{figure}

\subsubsection{HiMamba}
Fig.~\ref{TRIS-HAR-Arch} illustrates the HiMamba model, an extension of the standard Mamba model designed for 1-D sequences, adapted to handle CSI amplitude data of the multivariate time series form. This adaptation includes three main enhancements: embedding, a dual-stream framework, and a feature fusion layer.

\textbf{Embedding.}
CSI amplitude data comprises multiple frequency channels, each representing a univariate time series. These frequency tokens are linearly projected into frequency embeddings. For temporal channels, position embeddings are added to the non-linearly transformed time series data to capture temporal information. Formally, let $\boldsymbol{X} \in \mathbb{R}^{B \times L \times M}$ represent the input sequences with batch size $B$. The embedding process is expressed as:
\begin{equation}
    \begin{aligned}
        \boldsymbol{E}_f & = E_f (\boldsymbol{X}), \\
        \boldsymbol{E}_t & = E_t (\boldsymbol{X})+ E_{pos} (\boldsymbol{X}),
    \end{aligned}
\end{equation}
where $\boldsymbol{E}_f$ and $\boldsymbol{E}_t$ are the output frequency and temporal embeddings, respectively, each with dimensions $B \times L \times D$. Here, $E_f$, $E_t$ and $E_{pos}$ denote the frequency, temporal, and position embedding layers.

\textbf{Dual-stream framework.}
To effectively capture both temporal and frequency information in multivariate time series, we employ a dual-stream framework. This approach utilizes separate encoders in each stream to handle correlations related to time steps and frequency channels. The encoder structure is shown in Fig.~\ref{MambaEncoder}. Embedded tokens, $\boldsymbol{E}_f$ and $\boldsymbol{E}_t$, are processed through a sequence of Mamba blocks $L_i (i=1,\dots,N)$ within the encoder, producing the output $\boldsymbol{R}_N$. Each block integrates layer normalization \cite{ba2016layer}, depthwise separable convolution \cite{chollet2017xception}, and residual connections \cite{he2016deep}:
\begin{equation}
    \begin{aligned}
        \boldsymbol{H}_{n, 1} & = \rm{SSM}(\sigma(\rm{DW}(  \underbrace{\rm{MLP}(\rm{LN}(\boldsymbol{R}_{n-1}))}_{\boldsymbol{L}_{n, 1}} ))), \\
        \boldsymbol{H}_{n, 2} & = \sigma(  \underbrace{\rm{MLP}(\rm{LN}(\boldsymbol{R}_{n-1}))}_{\boldsymbol{L}_{n, 2}} ), \\
        \boldsymbol{R}_n & = \alpha \times \underbrace{\rm{MLP}(\boldsymbol{H}_{n, 1} \times \boldsymbol{H}_{n, 2})}_{\boldsymbol{F}_{n}} + \boldsymbol{R}_{n-1},
    \end{aligned}
\end{equation}
where {$\boldsymbol{R}_n \in \mathbb{R}^{B \times L \times D}$ is the output of the $n$-th block}, initialized as $\boldsymbol{R}_0 = \boldsymbol{E}$. In each block, $\rm{LN}(\boldsymbol{R}_{n-1})$ is linearly projected and split into $\boldsymbol{L}_{n, 1}$ and $\boldsymbol{L}_{n, 2}$. The abbreviations $\rm{DW}$, $\rm{LN}$, $\rm{MLP}$, and $\rm{SSM}$ stand for depthwise separable convolution, layer normalization, multi-layer perceptron, and the S6 model throughout~\eqref{SSMref}, respectively. $\sigma$ is the SiLU activation function \cite{hendrycks2016gaussian}, and $\alpha$ is a trainable parameter that adjusts the transformation within the layer $\boldsymbol{F}_n$ \cite{bachlechner2021rezero}.

The encoder's output token, $\boldsymbol{R}_N$, is passed through an average pooling layer to generate $\boldsymbol{P}_f \in \mathbb{R}^{B \times D}$ or $\boldsymbol{P}_t \in \mathbb{R}^{B \times D}$. These layers play a crucial role in summarizing key information across time and frequency dimensions, which is essential for robust feature extraction of subsequent classification tasks.

\begin{figure}[tbp]
    \centerline{\includegraphics[width=0.8\columnwidth]{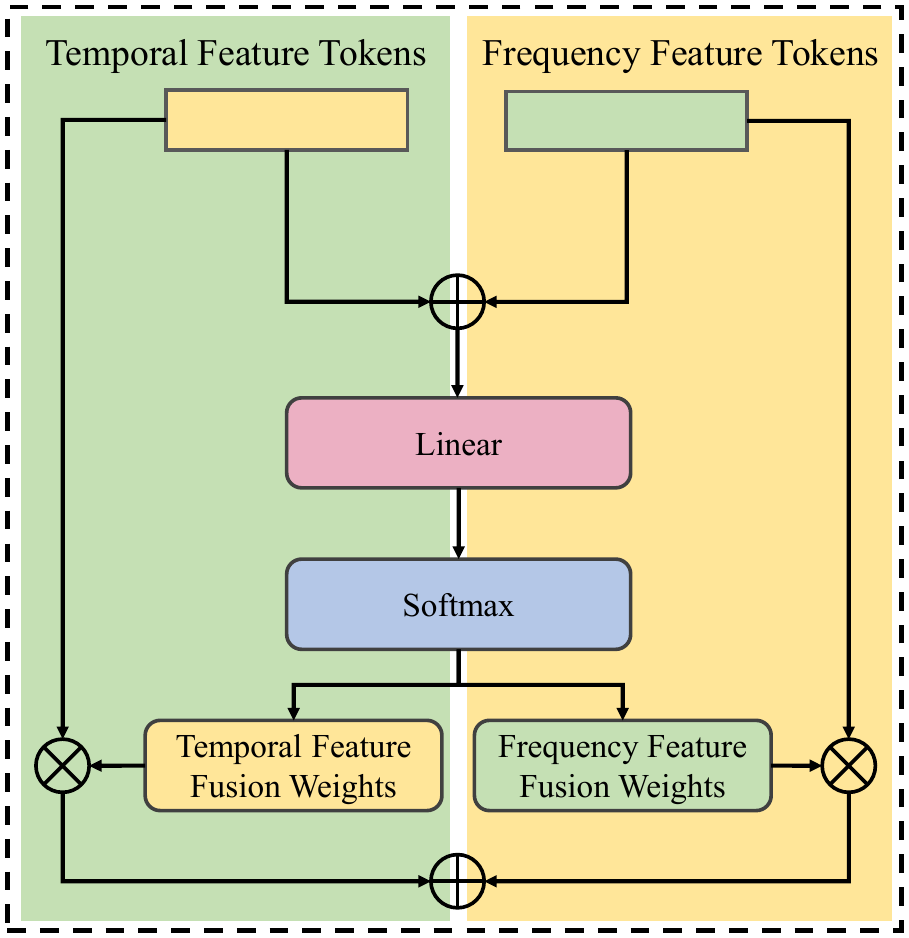}}
    \caption{The structure of fusion branch.}
    \label{Fusionbranch}
\end{figure}

\textbf{Feature fusion layer.}
To integrate features from both streams while effectively encoding step-wise and channel-wise correlations, a fusion mechanism is employed rather than simply concatenating all features. The structure of the fusion branch is shown in Fig.~\ref{Fusionbranch}. {Once the outputs of each stream are generated, they are concatenated and passed through a linear projection layer to produce $\boldsymbol{h} \in \mathbb{R}^{B \times 2D}$. A softmax function is then applied to compute the fusion weights, $\boldsymbol{f}_f \in \mathbb{R}^{B \times 1}$ and $\boldsymbol{f}_t \in \mathbb{R}^{B \times 1}$.} These weights are used to adjust the corresponding stream outputs, producing the final feature vector:
\begin{equation}
    \begin{aligned}
        \boldsymbol{h} & = \rm{LN}(\boldsymbol{P}_f \oplus \boldsymbol{P}_t), \\
        \boldsymbol{f}_f, \boldsymbol{f}_t & = \rm{Softmax}(\boldsymbol{h}), \\
        \boldsymbol{y} & = \boldsymbol{f}_f \odot \boldsymbol{P}_f \oplus \boldsymbol{f}_t \odot \boldsymbol{P}_t,
    \end{aligned}
\end{equation}
where $\oplus$ denotes the concatenation, and $\odot$ represents the Hadamard product. {The final output, $\boldsymbol{y} \in \mathbb{R}^{B \times n_c}$, is passed through a MLP head to obtain the final prediction $\hat{p}$:}
\begin{equation}
    \hat{p} = \rm{MLP}(\boldsymbol{y}),
\end{equation}
where $n_c$ is the number of classes in the task.

\section{Experiments and Performance Evaluation}
In this section, we evaluate the performance of our method using two publicly available datasets, UT-HAR and NTU-Fi, along with three real-world datasets: Laboratory, Office, and Meeting Room for comprehensive validation. All experiments are conducted on a server equipped with an AMD EPYC 7452 32-Core CPU @ 2.35 GHz, 256 GB RAM, and an NVIDIA GeForce RTX 3090, running on Linux and utilizing Python 3.9.19 and PyTorch 2.2.2.

\begin{table*}[htbp]
    \centering
    \caption{Evaluations of deep neural networks on two public datasets. (\textbf{Bold}: best; \underline{Underline}: 2nd best.)}
    \label{publicdataset}
    \setlength{\tabcolsep}{3.0mm}
    \begin{tabular}{cccccccc}
    \toprule
    \multirow{2}{*}{Models} & Datasets                             & \multicolumn{3}{c}{UT-HAR}                                   & \multicolumn{3}{c}{NTU-Fi}       \\
     & Methods                                                     & Acc (\%) & MACs (G)       & Params (M)                       & Acc (\%) & MACs (G)          & Params (M) \\
    \midrule
    \multirow{3}{*}{CNN-based} & CNN \cite{lecun1998gradient}      & 97.09    & \textbf{0.339} &  1.034                           &  98.11   &  \textbf{1.112}   &  2.106     \\
    & ResNet-18 \cite{he2016deep}                                  & 97.99    & 53.756         &  11.174                          &  97.73   &  196.054          &  11.174    \\
    & ConvNeXt \cite{liu2022convnet}                               & 98.19    & 26.756         &  0.844                           &  98.48   &  20.603           &  1.295     \\
    \multirow{2}{*}{LSTM-based} & LSTM \cite{hochreiter1997long}   & 90.86    & 1.819          &  \textbf{0.114}                           &  92.05   &  3.883          &  \textbf{0.242}  \\
    & MLSTMFCN \cite{karim2019multivariate}                        & 94.78    & 1.826  & 0.449                                    &  95.45   &  7.773            &  0.707 \\
    \multirow{2}{*}{GRU-based} & Bi-GRU-Attention \cite{meng2021wihgr} & 96.18    & 10.118     &  6.609                           &  96.59   &  26.882           &  12.754     \\
    & CNN-GRU \cite{huang2024tshnn}                                & 96.79    & 11.046         &  2.611                           &  96.97   &  26.762           &  12.688    \\
    \multirow{2}{*}{Transformer-based} & Encoder-only Transformer \cite{vaswani2017attention}  & \underline{98.29} & 101.229 &  12.661    &  98.11   &  204.522  &  12.789    \\
    & THAT \cite{li2021two}                                        & 98.09    & 30.364        &  7.252                            &  \underline{98.86}   &  76.210   &  8.219    \\
    \multirow{2}{*}{Mamba-based} & Mamba \cite{gu2023mamba}        & 97.19    & 2.890          &  0.363          &  98.26   &  6.275            &  0.396     \\
    & HiMamba (ours)                                               & \textbf{98.39} & \underline{1.176} & \underline{0.226}       & \textbf{99.24} & \underline{3.384} & \underline{0.258} \\
    \bottomrule
    \end{tabular}
\end{table*}

\subsection{Dataset Description}
\textbf{UT-HAR} \cite{yousefi2017survey}: The UT-HAR dataset, the first publicly available CSI dataset for human activity recognition, consists of seven activity categories performed by six participants. These activities include lying down, falling, walking, picking up, running, sitting down, and standing up, each performed in twenty trials. Data collection occurs in an indoor office environment using the Intel 5300 network interface card, equipped with three pairs of antennas and recording 30 subcarriers per pair. Each activity is sampled at 100 packets per second over a 20-second interval, resulting in a data structure of $30 \times 3 \times 2000$ for each CSI amplitude sequence. Following the approach of previous studies \cite{li2021two}, data segmentation is performed using a sliding window of 250, yielding approximately 5000 samples in total.

\textbf{NTU-Fi} \cite{yang2022efficientfi}: Collected with the Atheros CSI tool \cite{xie2015precise} on TP-Link N750 access points, one serving as the transmitter and the other as the receiver, the NTU-Fi dataset includes data from three antenna pairs operating at 5 GHz with a 40 MHz bandwidth. This setup allows for the extraction of 114 subcarriers of CSI data per timestamp. The dataset includes data on fourteen gait patterns and six activities, performed by seven female and thirteen male participants across three different environments. Each activity is performed twenty times by each participant, producing 400 samples per category, each recorded at a 500 Hz sampling rate over a one-second interval, resulting in a sample structure of $3 \times 114 \times 500$.

\subsection{Baselines}
We establish several baseline models, including CNN \cite{lecun1998gradient}, ResNet-18 \cite{he2016deep}, ConvNeXt \cite{liu2022convnet}, LSTM \cite{hochreiter1997long}, MLSTMFCN \cite{karim2019multivariate}, Bi-GRU-Attention \cite{meng2021wihgr}, CNN-GRU \cite{huang2024tshnn}, Encoder-only Transformer \cite{vaswani2017attention}, THAT \cite{li2021two}, and Mamba \cite{gu2023mamba}.

\subsection{The Effectiveness and Efficiency of Proposed HiMamba}
We evaluate the baseline models based on three criteria: accuracy (Acc), which measures classification efficacy; multiply-accumulate operations (MACs), which evaluate computational complexity; and the number of parameters (Params), indicating GPU memory requirements. We use the Adam optimizer with a learning rate of $10^{-4}$, a batch size of 32, and 40 training epochs.

\subsubsection{Effectiveness}
Table~\ref{publicdataset} presents the accuracy of HiMamba and various baselines on the UT-HAR and NTU-Fi datasets. Using the NTU-Fi public dataset as an example, the following observations can be made:

\begin{itemize}
    \item [\textit{i})] 
    Our HiMamba model demonstrates high effectiveness. The average accuracy of the proposed HiMamba approach is 99.24\%, outperforming other baselines. Compared to the LSTM-based and GRU-based models, HiMamba surpasses them by more than two percentage points. Despite THAT being a larger model with approximately 30 times more parameters, HiMamba still outperforms it by nearly one percentage point. This result can be attributed to HiMamba's more efficient architecture and superior ability to capture relevant features with fewer parameters, resulting in better generalization and performance.
    \item [\textit{ii})]
    The Transformer-based model, THAT, performs better than the CNN-based, LSTM-based, and GRU-based models, primarily due to its strong expressiveness. Despite this strong baseline, our HiMamba model achieves nearly a one percentage point improvement while using only 3.14\% of THAT's parameters, demonstrating the high efficacy of our design.
\end{itemize}

The trend on the UT-HAR dataset is quite similar. The proposed HiMamba model achieves an accuracy of 98.39\%, which is also the highest among all models.

\subsubsection{Efficiency} 
The model complexity comparison for all methods is listed in Table~\ref{publicdataset}, with data collected using the THOP tool\footnote{https://github.com/Lyken17/pytorch-OpCounter}. The CNN has the lowest computational complexity at 1.112 G, and the LSTM has the lowest number of parameters at 0.242 million. The Encoder-only Transformer has the highest computational complexity at 204.522 G and 12.789 million parameters. Additionally, the THAT method has a computational complexity of 76.210 G and 8.219 million parameters. Compared to THAT, our HiMamba model exhibits significantly lower model complexity in both the number of parameters and computational complexity. HiMamba has only 0.258 million parameters, which is just 3.14\% of THAT's parameters. HiMamba's computational complexity is 3.384 G, compared to THAT's 76.210 G, being $22.5\times$ less complex in terms of MACs. This demonstrates that HiMamba is an effective and lightweight model, achieving relatively low computational complexity while maintaining a reasonable number of parameters.

In summary, the HiMamba model strikes an optimal balance between accuracy, model size, and computational complexity, establishing it as a highly efficient tool that achieves state-of-the-art performance. Notably, it is substantially lighter than comparable models. The superior performance of HiMamba can be attributed to two key innovations. First, its dual-stream architecture adeptly captures both temporal and frequency features, significantly outperforming traditional methods by modeling complex dependencies in the data more effectively. This dual processing pathway greatly enhances the model's ability to extract relevant information from diverse signal patterns, thereby boosting its accuracy. Second, HiMamba incorporates a highly optimized parameterization scheme that reduces computational costs by focusing on the most relevant input data. This approach contrasts sharply with more complex models like Transformers, which uniformly process all inputs. As a result, HiMamba not only achieves excellent performance but also maintains fewer parameters and lower computational complexity, thereby increasing its robustness and efficiency.




\subsection{TRIS-enabled HAR}
We conduct experiments in the Laboratory dataset to assess the impact of TRIS on HAR performance. The setup simulates a typical indoor environment with common obstacles and a TRIS, as shown in Fig.~\ref{F4}~\subref{F4-a}. We evaluate the system's ability to recognize six human activities: kicking, picking up, sitting down, standing, standing up, and walking, in both RIS-enabled and non-RIS scenarios. Performance metrics such as accuracy, precision, recall, and F1-score are used to quantify improvements.

\begin{table}[tbp]
    \centering
    \caption{Performance metrics of TRIS-HAR system in laboratory settings with and without (w/o) the TRIS.}
    \label{RISEnabled}
    \begin{tabular}{ccccc}
    \toprule
    Environment  & Accuracy & Precision & Recall & F1-score \\
    \midrule
    Laboratory - w/o TRIS   & 0.8500    & 0.8551     & 0.8475  & 0.8476    \\
    Laboratory - with TRIS  & 0.9806    & 0.9806     & 0.9807  & 0.9806    \\
    \bottomrule   
    \end{tabular}
\end{table}

We validate the HiMamba model on the Laboratory datasets (TRIS-enabled and non-TRIS scenarios) and adopt the Adam optimizer with a learning rate of $10^{-3}$, batch size of 32, and 50 training epochs. Table~\ref{RISEnabled} shows the performance metrics for the TRIS-HAR system with and without the TRIS. The TRIS-enabled setup significantly outperforms the non-TRIS setup across all metrics. Accuracy increases from 85.00\% to 98.06\%, demonstrating a substantial improvement.

\begin{figure}[htbp]
    \centerline{\includegraphics[width=1.0\columnwidth]{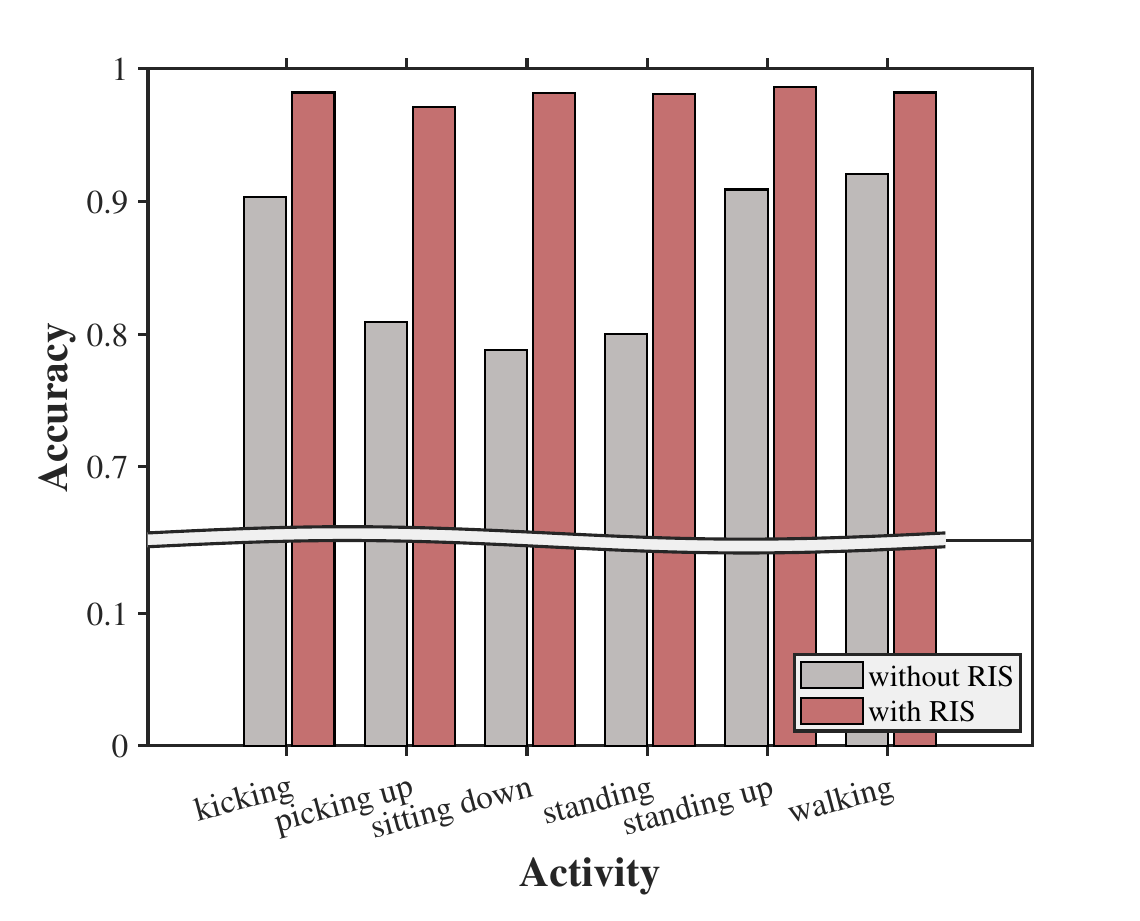}}
    \caption{Recognition accuracy for different human activities in the laboratory environment with and without the TRIS.}
    \label{DifferentAction}
\end{figure}

Fig.~\ref{DifferentAction} illustrates the human activity recognition accuracy in the laboratory. The TRIS-HAR system achieves an average accuracy of 98.06\% across six human activities. Specifically, the TRIS-enabled recognition accuracy for kicking, picking up, sitting down, standing, standing up, and walking is 98.21\%, 97.14\%, 98.15\%, 98.08\%, 98.61\%, and 98.21\%, respectively, with even the lowest accuracy above 97\%. The experimental results demonstrate that the TRIS-enabled HAR system achieves high recognition accuracy for various human activities.

\subsection{Robustness}
In this subsection, we evaluate the robustness of the TRIS-HAR system across different environmental settings. Robustness is crucial to ensure that the system maintains high performance under various real-world conditions, which may include changes in the environment and the presence of interference.

We test the TRIS-HAR system in two different environments: an office and a meeting room. These environments are chosen to represent typical indoor settings with varying characteristics. Each environment is equipped with common obstacles and potential interference sources to access the system's robustness. The experiments involve a set of predefined human activities, performed under the same protocol as in previous experiments, ensuring consistency in the evaluation.

\begin{table}[tbp]
    \centering
    \caption{Performance metrics of TRIS-HAR system in different environments.}
    \label{RobustnessTest}
    \setlength{\tabcolsep}{3mm}
    \begin{tabular}{ccccc}
    \toprule
    Environment  & Accuracy & Precision & Recall & F1-score \\
    \midrule
    Office       & 0.9833   & 0.9832    & 0.9837 & 0.9834    \\
    Meeting room & 0.9778   & 0.9764    & 0.9771 & 0.9766    \\
    \bottomrule   
    \end{tabular}
\end{table}

Table~\ref{RobustnessTest} presents the performance metrics of the TRIS-HAR system in both the office and meeting room environments. The system demonstrates high accuracy and robust performance across both settings, showing minimal degradation despite environmental changes.

The TRIS-HAR system consistently achieves high performance metrics in both environments. In the office setting, the system achieves an accuracy of 98.33\%, a precision of 98.32\%, a recall of 98.37\%, and an F1-score of 98.34\%. In the meeting room, the system achieves an accuracy of 97.78\%, a precision of 97.64\%, a recall of 97.71\%, and an F1-score of 97.66\%. The results indicate that the TRIS-HAR system is not significantly affected by environmental changes or obstacles and maintains its ability to accurately recognize various human activities. This high level of performance across different settings underscores the system's robustness and its potential applicability in diverse indoor environments.

\subsection{Ablation Test}
{We conduct an ablation study to evaluate the contributions of the fusion mechanism and the dual-stream structure. The results are detailed in Table~\ref{ablationstudy}. In each test, we remove a specific component from the full model to analyze its impact on the overall performance.}

{From Table~\ref{ablationstudy}, it is evident that eliminating the fusion mechanism and substituting it with simple concatenation results in notable accuracy declines. Specifically, this change leads to a decrease of 3.03\% and 2.41\% on the NTU-Fi and UT-HAR datasets, respectively, with a more pronounced drop of 5.00\% in complex environments such as the meeting room. These findings confirm the effectiveness of the weighted fusion mechanism in integrating information from both streams, thereby enhancing overall recognition performance.}

{Subsequently, we assess the effects of removing either the temporal or frequency module while retaining the other. Retaining the temporal module and removing the frequency module results in a decrease of 4.83\% on UT-HAR and 4.15\% on NTU-Fi, underscoring the supportive role of frequency-domain features in the model's interpretation of signal variations, even though temporal features are more pivotal.}

{Conversely, retaining the frequency module while removing the temporal module leads to more substantial performance declines, with decreases of 9.43\% on UT-HAR and 8.18\% on NTU-Fi. This indicates that temporal features are crucial for activity recognition, particularly in complex settings such as the meeting room, where the accuracy dropped by 12.45\%. Temporal features appear more intuitive and essential in capturing human motion sequences.}

{In conclusion, the results of the ablation study clearly demonstrate that each component, including the fusion mechanism, temporal stream, and frequency stream, plays an indispensable role in achieving the high accuracy of HiMamba. Specifically, the dual-stream structure effectively complements both temporal and frequency features to maximize recognition accuracy.}

\begin{table*}[tbp]
    \centering
    \caption{{Ablation study results compared with the full HiMamba model. The accuracy is shown in percentage (\%).}}
    \label{ablationstudy}
    \setlength{\tabcolsep}{2.8mm}
    \begin{tabular}{lcccccccccc}
    \toprule
    \multirow{2}{*}{Model} & \multicolumn{2}{c}{UT-HAR} & \multicolumn{2}{c}{NTU-Fi} & \multicolumn{2}{c}{Laboratory} & \multicolumn{2}{c}{Office}  & \multicolumn{2}{c}{Meeting room}     \\
    & Accuracy & $\Delta$  & Accuracy & $\Delta$  & Accuracy & $\Delta$   & Accuracy & $\Delta$  & Accuracy & $\Delta$  \\
    \midrule
    HiMamba                & 98.39 & -     & 99.24 & -     & 98.06 & -     & 98.33 & -     & 97.78 & -  \\
    - Concatenation Fusion & 95.98 & -2.41 & 96.21 & -3.03 & 93.61 & -4.45 & 96.50 & -1.83 & 92.78 & -5.00  \\
    - Temporal Module      & 93.56 & -4.83 & 95.09 & -4.15 & 91.67 & -6.39 & 94.17 & -4.16 & 90.50 & -7.28  \\
    - Frequency Module     & 88.96 & -9.43 & 91.06 & -8.18 & 86.67 & -11.39 & 92.78 & -5.55 & 85.33 & -12.45  \\
    \bottomrule
    \end{tabular}
\end{table*}

\subsection{Compared with Existing Methods}

Table~\ref{ComparedwithWorks} presents the recognition accuracy and the number of parameters of the proposed HiMamba model compared with other baselines across the Laboratory, Office, and Meeting room datasets. HiMamba consistently outperforms the other models in terms of accuracy on all three datasets. Specifically, on the Laboratory dataset, HiMamba is 1.95\% and 2.23\% more accurate than ConvNeXt and Bi-GRU-Attention, respectively. On the Office dataset, HiMamba surpasses ConvNeXt and Bi-GRU-Attention by 2.16\% and 1.50\%, respectively. In the Meeting room dataset, HiMamba outperforms ConvNeXt and Bi-GRU-Attention by 4.78\% and 5.61\%, respectively.

Compared to the Transformer-based model THAT, our model achieves marginally better accuracy by 0.56\%, 0.5\%, and 0.84\% across different datasets. However, THAT has a significantly larger model size, with approximately 14 times more parameters than ours. In contrast, when compared to the LSTM model, HiMamba, which has a higher number of parameters at 0.737 million, demonstrates substantial improvements in accuracy, achieving 7.23\%, 6.00\%, and 6.61\% higher accuracy across the three datasets. This indicates that HiMamba successfully combines low model complexity with high performance.

\begin{table*}[tbp]
    \centering
    \caption{Performance comparison of the proposed approach with other existing methods. (\textbf{Bold}: best; \underline{Underline}: 2nd best.)}
    \label{ComparedwithWorks}
    \setlength{\tabcolsep}{3.0mm}
    \begin{tabular}{cccccccc}
    \toprule
    \multirow{2}{*}{Models} & Datasets & \multicolumn{2}{c}{Laboratory} & \multicolumn{2}{c}{Office}  & \multicolumn{2}{c}{Meeting room}     \\
     & Methods  & Acc (\%) & Params (M)   & Acc (\%)   & Params (M)  & Acc (\%)          & Params (M)  \\
    \midrule
    \multirow{3}{*}{CNN-based} & CNN \cite{lecun1998gradient} & 96.11 & 3.958 & 96.50 & 3.958 & 94.67 & 3.958  \\
    & ResNet-18 \cite{he2016deep} & 95.28 & 11.173 & 96.50 & 11.173 & 95.67 & 11.173    \\
    & ConvNeXt \cite{liu2022convnet} & 96.11 & 0.951 & 96.17 & 0.951 & 93.00 & 0.951     \\
    \multirow{2}{*}{LSTM-based} & LSTM \cite{hochreiter1997long} & 90.83 & \textbf{0.144} & 92.33 & \textbf{0.144} & 91.17 & \textbf{0.144} \\
    & MLSTMFCN \cite{karim2019multivariate} & 94.44 & 8.730 & 96.50 & 8.730 & 94.67 & 8.730 \\
    \multirow{2}{*}{GRU-based} & Bi-GRU-Attention \cite{meng2021wihgr} & 95.83 & 4.151 & 96.83 & 4.151 & 92.17 & 4.151 \\
    & CNN-GRU \cite{huang2024tshnn} & 96.39 & 4.085 & 96.83 & 4.085 & 95.83 & 4.085    \\
    \multirow{2}{*}{Transformer-based} & Encoder-only Transformer \cite{vaswani2017attention} & 96.94 & 16.801 & 96.17 & 16.801 & 94.33 & 16.801 \\
    & THAT \cite{li2021two} & \underline{97.50} & 10.746 & \underline{97.83} & 10.746 & \underline{96.94} & 10.746 \\
    \multirow{2}{*}{Mamba-based} & Mamba \cite{gu2023mamba} & 95.28 & 1.398 & 93.83 & 1.398 & 91.94 & 1.398 \\
    & HiMamba (ours) & \textbf{98.06} & \underline{0.737} & \textbf{98.33} & \underline{0.737} & \textbf{97.78} & \underline{0.737} \\
    \bottomrule
    \end{tabular}
\end{table*}

\section{Conclusion}
In this paper, we presented the TRIS-HAR system, which leverages transmissive reconfigurable intelligent surfaces to enhance RF-based human activity recognition. The system demonstrated significant improvements in HAR accuracy, increasing from 85.00\% to 98.06\% in laboratory settings, while maintaining high performance in office and meeting room environments. By generating multiple independent signal paths, the TRIS-HAR system effectively mitigated multipath fading and improved signal clarity. Additionally, the HiMamba model, specifically designed to process the enhanced CSI data, successfully captured both temporal and frequency-based features. This model not only achieved high recognition accuracy across public datasets but also maintained computational efficiency. Our results emphasize the potential of integrating TRIS technology with advanced modeling techniques to improve HAR, particularly in complex indoor environments.

Recent advancements in stacked intelligent metasurfaces underscore the potential to further enhance wireless sensing and signal processing capabilities by stacking multiple transmissive metasurface layers. These SIMs can execute complex tasks, such as MIMO precoding and DOA estimation directly in the wave domain, markedly reducing latency and energy consumption while enhancing computational efficiency. Extending our TRIS-HAR system to incorporate SIM technology could improve human activity recognition by enabling more precise and adaptive signal manipulation. This prospective direction holds promise for advancing the accuracy and robustness of HAR systems in complex indoor environments, as well as opening opportunities for advanced applications such as smart home automation and remote disaster monitoring.

\bibliographystyle{IEEEtran}
\bibliography{revise_v1.bib}

\newpage

\vfill

\end{document}